\pdfoutput=1
\documentclass[iop,twocolumn,apj]{emulateapj}

\usepackage{graphicx}
\usepackage{amsmath}
\usepackage{multirow}
\newcommand       \be		{\begin{equation}}
\newcommand       \ee		{\end{equation}}

\newcommand       \erg		{\,{\rm erg}}
\newcommand       \K		{\,{\rm K}}

\newcommand       \cm		{\,{\rm cm}}

\newcommand       \g		{\,{\rm g}}

\newcommand       \rhocgs       {{\rm g/cm}^{3}}

\begin{document}
\title{Make Super-Earths, Not Jupiters: \\
Accreting Nebular Gas onto Solid Cores at 0.1 AU and Beyond}
\author{Eve J. Lee\altaffilmark{1}, Eugene Chiang\altaffilmark{1,2},
  Chris W. Ormel\altaffilmark{1,3}} 
\altaffiltext{1}{Department of Astronomy, University of California Berkeley, Berkeley, CA 94720-3411, USA; 
evelee@berkeley.edu, echiang@astro.berkeley.edu, ormel@berkeley.edu}
\altaffiltext{2}{Department of Earth and Planetary Science, University of California Berkeley, Berkeley, CA 94720-4767, USA}
\altaffiltext{3}{Hubble Fellow}

\begin{abstract} 
  Close-in super-Earths having radii 1--4 $R_\oplus$ may possess
  hydrogen atmospheres comprising a few percent by mass of their rocky
  cores.  We determine the conditions under which such atmospheres can
  be accreted by cores from their parent circumstellar disks.
  Accretion from the nebula is problematic because it is too
  efficient: we find that 10~$M_\oplus$
  cores embedded in solar metallicity disks tend to undergo runaway
  gas accretion and explode into Jupiters, irrespective of orbital
  location.  The threat of runaway is especially dire at $\sim$0.1 AU,
  where solids may coagulate on timescales orders of magnitude shorter
  than gas clearing times; thus nascent atmospheres on close-in orbits
  are unlikely to be supported against collapse by planetesimal
  accretion. The time to runaway accretion is well approximated by the
  cooling time of the atmosphere's innermost convective zone, whose
  extent is controlled by where H$_2$ dissociates.
  Insofar as the temperatures characterizing H$_2$ dissociation are
  universal, timescales for core instability tend not to vary with
  orbital distance --- and to be alarmingly short for 10~$M_\oplus$
  cores.  Nevertheless, in the thicket of parameter space, we identify 
  two scenarios, not mutually exclusive, that can reproduce the
  preponderance of percent-by-mass atmospheres for super-Earths at
  $\sim$0.1 AU, while still ensuring the formation of Jupiters at
  $\gtrsim 1$ AU. Scenario (a): planets form in disks with dust-to-gas
  ratios that range from $\sim$20$\times$ solar at 0.1 AU to
  $\sim$2$\times$ solar at 5 AU. Scenario (b): the final assembly of
  super-Earth cores from mergers of proto-cores --- a process that
  completes quickly at $\sim$0.1 AU once begun --- is delayed by gas
  dynamical friction until just before disk gas dissipates completely.
  Both scenarios predict that the occurrence rate for
    super-Earths vs.~orbital distance, and the corresponding rate for
    Jupiters, should trend in opposite directions, as the former
    population is transformed into the latter: as gas giants become
    more frequent from $\sim$1 to 10 AU,
    super-Earths should become more rare.
\end{abstract}

\section{INTRODUCTION}
\label{sec:intro}
Core-nucleated instability is a widely-believed mechanism by which gas
giant planets form (see \citealt{2000prpl.conf.1081W} and
\citealt{2007prpl.conf..591L} for reviews).  The theory states that a
solid core of rock and metal, when embedded in a gas-rich nebula,
undergoes ``runaway gas accretion'' to become a gas giant like
Jupiter, if the core mass is sufficiently large
\citep{1974Icar...22..416P, 1978LPI.....9..459H, 1978PThPh..60..699M,
  1980PThPh..64..544M, 1982P&SS...30..755S}.  In static models in
  which the planet's nascent gas envelope is powered by steady accretion of
rocky planetesimals \citep[e.g.,][]{1980PThPh..64..544M,
  1982P&SS...30..755S, 2006ApJ...648..666R, 2011ApJ...727...86R},
runaway accretion occurs at a ``critical core mass'' above which the
envelope fails to maintain hydrostatic equilibrium.
Identifying the critical core mass with
hydrostatic disequilibrium
is specific to static models. In time-dependent
models \citep[e.g.,][]{1996Icar..124...62P}, runaway accretion is
characterized by envelope masses that grow superlinearly with time.
Physically, runaway is triggered once the self-gravity of the
atmosphere becomes significant, i.e., when the envelope has about as
much mass as the core. The critical core mass in time-dependent models 
is that for which runaway accretion occurs just within the gas disk lifetime; 
it is less related to hydrostatic disequilibrium than to increasing thermal 
disequilibrium: runaway accretion occurs because of runaway cooling.

Critical core masses are typically quoted to be $\sim$$10 M_{\oplus}$
\citep[e.g.,][]{1980PThPh..64..544M, 1982P&SS...30..755S,
  1996Icar..124...62P, 2000ApJ...537.1013I}.  As long as the planet's
envelope is connected to the nebula by a radiative outer layer, the
critical core mass does not depend much on nebular conditions
\citep{1980PThPh..64..544M, 1982P&SS...30..755S}.  The radiative
zone's steep rise in density and pressure tends to decouple the
planet's interior from the external environment.  Thus the theoretical
prejudice is that $10$-$M_\oplus$ cores nucleate gas giants whether
they are located at $\sim$0.1 AU or $\sim$5 AU.
\citet{1982P&SS...30..755S} justifies analytically how $10 M_\oplus$
is a characteristic critical core mass, under the assumption that most
of the envelope mass is radiative and has constant opacity.\footnote{
 If the envelope is more nearly adiabatic,
the critical
core mass depends more sensitively on nebular boundary conditions
(e.g., \citealt{wuchterl93} and \citealt{2000ApJ...537.1013I}). 
We will find that the envelopes of {\it Kepler} super-Earths 
have substantial outer radiative zones.}

The discovery of {\it Kepler} super-Earths presents a seeming crisis
for the core instability theory. About 1 in 5 Sun-like stars harbor
planets having radii of 1--4 $R_{\oplus}$ at distances of 0.05--0.3 AU
\citep{2010Sci...330..653H, 2013ApJS..204...24B, 2013ApJ...770...69P,
  2013ApJ...778...53D, 2013ApJ...766...81F, 2014ApJ...784...45R}.
Transit-timing analyses \citep{2013ApJ...772...74W} and
radial-velocity measurements \citep{2014ApJ...783L...6W} establish
these super-Earths to have masses of 2--20 $M_\oplus$ --- these are in
the range of critical core masses cited for runaway gas accretion.
Yet such super-Earths apparently retain only small amounts of gas:
only $\sim$3--10\% by mass \citep[e.g.,][]{2014ApJ...792....1L} or
even less \citep[e.g.,][]{2010ApJ...712..974R, 2010ApJ...716.1208R},
based on modeling of observed radius-mass data. The prevalence of
super-Earths is consonant with the rarity of Jupiters at these
distances \citep[e.g.,][]{jones03, udry03, 2012ApJ...753..160W,
  2013ApJ...766...81F, 2013ApJ...767L..24D}.

How did super-Earths having masses of $\sim$10 $M_\oplus$ avoid
becoming gas giants? How did they acquire the modest gas envelopes
that they are inferred to have?  We begin by offering some general
perspectives on in-situ formation at small orbital distances that will
dictate our approach to answering these questions.\footnote{
One may question our assumption of in-situ formation
and our suppression of migration.
Types I and II migration are infamously rapid,
especially at the small stellocentric distances where super-Earths currently
reside. Nevertheless, co-rotation torques render
both the timescale and even direction of migration
uncertain (see \citealt{2012ARA&A..50..211K} for a review).
Whereas what halts migration remains unclear,
simulations of in-situ formation 
can reproduce the
observed distributions of orbital periods,
planetary sizes, and mutual inclinations,
using only a modicum of input parameters \citep{2013ApJ...775...53H}. 
And whether or not super-Earth cores migrated inward, it remains 
to be explained how they acquired their atmospheres and avoided runaway: 
this is the problem we address in this paper, and elements of our solution 
should be independent of the manner by which cores are emplaced.
}
As we show below,
the problem of avoiding runaway gas accretion is especially severe
for close-in super-Earth cores because they are not expected to have 
significant sources of power that can keep their atmospheres in strict
equilibrium. Because their atmospheres are free to cool and contract,
they are at great risk of accreting large amounts of gas from the
nebula and exploding into Jupiters.

\subsection{For Super-Earths, Accretion of Solids Completes Before Accretion of Gas}
\label{ssec:battery}
Many of the core instability studies cited above assume 
that solid cores accrete rocks and gas simultaneously.
Gravitational energy released from solids raining down upon
the core heats the envelope and acts as a 
battery: the luminosity $L_{\rm acc}$ derived from
planetesimal accretion is analogous to nuclear
power in stars. In static models, $L_{\rm acc}$ is a prescribed
constant in time. For example, \citet{2006ApJ...648..666R}, using
prescriptions for $L_{\rm acc}$ that 
depend on orbital distance, finds that critical core masses $M_{\rm crit}$ range from
$\sim$0.1--100 $M_\oplus$ over $\sim$0.05--100 AU.\footnote{
\citet[][their Figure 7]{2006ApJ...648..666R}
reports $M_{\rm crit} \sim 7 M_{\oplus}$ at 0.1 AU
--- which taken at face value violates the observation that
super-Earths having just these masses, and not Jupiters,
prevail at such distances.
But this estimate of $M_{\rm crit}$
assumes that the planet's envelope is bounded by the
Bondi radius $R_{\rm B}$ rather than the Hill radius $R_{\rm H}$.
At orbital distances $a \sim 0.1$ AU, the opposite is likely to be true
(see, e.g., our \S\ref{sssec:BC}). 
For the strictly adiabatic envelopes assumed by \citet{2006ApJ...648..666R} at $\sim$0.1 AU,
replacing his $R_{\rm B}$ with the smaller $R_{\rm H}$
implies that 7-$M_\oplus$ cores
have only small gas-to-core ratios and would not run away.
More problematic is 
his assumption of strict adiabaticity which follows from an
unsustainably large planetesimal
accretion rate.
}
In time-dependent models, $L_{\rm acc}$ 
is a prescribed function of time that 
enables one to follow the envelope's quasi-static contraction and mass gain from the nebula
\citep{1996Icar..124...62P,
  2000Icar..143....2B, 2010Icar..209..616M, 2011ApJ...738...59R}.

At $\sim$5 AU and beyond, the usual practice of accounting for $L_{\rm
  acc} \neq 0$ is sensible.
Coagulation of solids can play out any number of ways, especially
at large orbital distances where to form planets
within a Hubble time, 
accretion of planetesimals is necessarily gravitationally focussed.
These gravitational focussing factors (a.k.a.~Safronov numbers)
are uncertain, depending sensitively on the size distribution
of planetesimals and the means by which velocity dispersions are damped
(see \citealt{2004ARA&A..42..549G} for a pedagogic review). 
Among the myriad planetesimal accretion histories $L_{\rm acc} (t)$ that
are imaginable outside a few AU, many have durations at least as long, if not 
longer, than gas disk lifetimes of several Myrs,
validating the conventional $L_{\rm acc}$-powered models for core instability.

But closer to the star, the universe of
possibilities narrows considerably, assuming that planets form in-situ 
\citep{2012ApJ...751..158H, 2013ApJ...775...53H, 2013MNRAS.431.3444C}.
Even without gravitational focussing, the time to coagulate a solid
core of mass $M_{\rm core}$ and radius $R_{\rm core}$ in-situ at
$\sim$0.1 AU is astonishingly short:
\begin{equation}
\begin{split}
t_{\rm coagulate} &\sim \frac{M_{\rm core}}{\dot{M}_{\rm core}} \sim \frac{M_{\rm core}}{\rho_{\rm s} R_{\rm core}^2 v_{\rm rel}} \\
&\sim \frac{M_{\rm core}}{(\Sigma_{\rm s}/H) R_{\rm core}^2 v_{\rm rel}} \sim \frac{M_{\rm core}}{\Sigma_{\rm s} R_{\rm core}^2 \Omega}\, ,
\end{split}
\label{eq:tcoag1}
\end{equation}
where $\rho_{\rm s}$ and $\Sigma_{\rm s}$ are the planetesimals'
volume density and surface density, respectively; $H$ is their scale
height; $v_{\rm rel} \sim H \Omega$ is their velocity dispersion,
assumed isotropic; and $\Omega$ is the local orbital frequency. If we
assume that the disk has the ``minimum mass'' needed to spawn a core
from an annulus of radius $a$ and width $\Delta a \sim a$ --- so that
$\Sigma_{\rm s} \sim M_{\rm core}/a^2$ --- then we arrive at a simple
expression for the coagulation time:
\begin{equation}
\begin{split}
t_{\rm coagulate} &\sim \left( \frac{a}{R_{\rm core}} \right)^2 \Omega^{-1} \\
&\sim 10^4 \, {\rm yr}\, \left(\frac{a}{0.1 \,{\rm AU}}\right)^{3.5} \left(\frac{1.6 R_{\oplus}}{R_{\rm core}}\right)^2 \, .
\end{split}
\label{eq:tcoag3}
\end{equation}
This is 2--3 orders of magnitude shorter than 
gas disk dissipation timescales of
$t_{\rm disk} \sim 0.5$--10 Myr 
\citep{2009AIPC.1158....3M, alexander14}.
The lesson here is that close-in orbits have compact areas $a^2$
(i.e., high surface densities) and short dynamical times $\Omega^{-1}$
(i.e., the local clock runs dizzyingly fast), effecting rapid
coagulation.  And equation (\ref{eq:tcoag3}) represents an upper limit 
--- both because we have neglected gravitational
focussing, and because we have assumed a minimum disk surface density.

The coagulation time $t_{\rm coagulate}$ characterizes the last doubling of mass of the core,
irrespective of whether that doubling is achieved by accreting
small planetesimals (``minor mergers'')
or by giant impacts between oligarchs (``major mergers''). 
Whatever dregs of planetesimals remain from the last doubling are consumed over timescales 
comparable to $t_{\rm coagulate}$.
Because $t_{\rm coagulate} \ll t_{\rm disk}$, the standard picture of
having planetesimals accrete contemporaneously with disk gas is not
appropriate for close-in super-Earths.  Solids finish accreting well
before gas finishes accreting --- indeed even before gas starts to
accrete in earnest. 

One consequence of $L_{\rm acc} = 0$ is that gas envelopes can freely cool,
contract, and accrete more gas: they are
especially vulnerable to runaway
gas accretion. This feature of in-situ formation only heightens the
mystery of why the {\it Kepler} spacecraft and ground-based radial
velocity surveys find an abundance of super-Earths but not Jupiters at
$\sim$0.1 AU.

\subsection{Mission and Plan for this Paper} 
\label{ssec:mission_impossible}
We seek here to unravel this mystery --- to understand how super-Earths
avoid runaway gas accretion, and just as importantly,
to see if we can reproduce their observationally inferred gas fractions.
Motivated by the considerations in \S\ref{ssec:battery},
we will set the planetesimal accretion rate to zero when we construct
models for how close-in super-Earth cores accrete gas from their natal
disks. Similar passively cooling atmospheres have been computed
at orbital distances of 5 AU 
\citep[e.g.,][]{2000ApJ...537.1013I, 2005A&A...433..247P} and beyond 
\citep[][]{2014ApJ...786...21P}.
In these models, cooling regulates the accretion of gas.
As the envelope radiates away its energy, it undergoes Kelvin-Helmholtz 
contraction and accretes more gas from the surrounding nebula.
Both \citet{2000ApJ...537.1013I} and \citet{2005A&A...433..247P} conclude
that Jupiter could have formed in-situ as long as $M_{\rm core} \gtrsim 5 M_{\oplus}$.
Distant extrasolar gas giants like those orbiting HR 8799 (\citealt{marois08};
\citealt{marois10}), located as far as $\sim$30--70 AU from their host star,
might also have formed via core instability,
starting from core masses as small as $\sim$$4 M_\oplus$
(\citealt{2014ApJ...786...21P}; but see \citealt{kratter10}
and references therein for alternative formation channels
involving gravitational instability or outward migration of solid
cores).

Our focus here is on the acquisition of gas envelopes 
at 0.1 AU. Pioneering studies at these close-in distances by
\citet{2012ApJ...753...66I} and \citet{2014ApJ...791..103B}
concentrate on the case of the multi-planet system orbiting
Kepler-11 \citep{lissauer11, lissauer13}. 
From Figure 2 of \citet[][]{2012ApJ...753...66I}, 
gas accretion onto cores is slowed within hotter and dustier disks,
but gas disk lifetimes of $t_{\rm disk} \sim 0.5$--10 Myr
\citep{2009AIPC.1158....3M, alexander14} 
are long enough that 10~$M_\oplus$ cores are still expected to run away.
We will confirm these results and chart new regions of parameter space
(exploring, e.g., supersolar metallicities)
to find accretion histories that do succeed
in circumventing runaway for 10~$M_\oplus$ cores.
Although \citet{2012ApJ...753...66I} set the planetesimal accretion
luminosity $L_{\rm acc} = 0$ (as we have explained is realistic),
their models still feature a large internal luminosity:
one that emanates from the rocky core with its finite heat capacity.
In \S\ref{sssec:checks}, we show by contrast
that this power input is actually negligible.
\citet{2014ApJ...791..103B} focus on
Kepler-11f, and find that a solid core of mass $M_{\rm core} \simeq 2
M_{\oplus}$ at 0.5 AU can safely avoid runaway,
accreting an atmosphere for which $M_{\rm gas}/M_{\rm core} \sim 2\%$, 
so long as the disk disperses in 2 Myr.  
By comparison, our study is more general and
will encompass the super-Earth population as a whole,
located between 0.05--0.5 AU. Most of their measured masses range from
5--10$M_\oplus$ \citep{2013ApJ...772...74W, 2014ApJ...783L...6W}

Because our model gas envelopes are not heated,
they are especially susceptible to runaway; i.e., our
calculated runaway times $t_{\rm run}$ are strict lower bounds.
Thus when we identify those conditions for which 
$t_{\rm run} > t_{\rm disk}$---thereby making the universe safe 
for super-Earths---such solutions should be robust.
We describe the construction of our model atmospheres in \S\ref{sec:model}.
Results for $t_{\rm run}$ and its variation with nebular conditions
and core masses are presented in \S\ref{sec:results}.
Readers interested only in our solution to how super-Earths remain
super-Earths may skip directly to \S\ref{sec:disc}.
There we propose two possible scenarios by which close-in planets
avoid runaway --- and also calculate their expected final gas contents.
A summary is given in \S\ref{sec:concl}. 

\section{TIME-DEPENDENT MODEL ATMOSPHERES}
\label{sec:model}
We model how a rocky core embedded in a gaseous circumstellar
disk accretes an atmosphere. The gas envelope is assumed
spherically symmetric: we solve for how density and temperature vary with 
radius and time, subject to outer boundary conditions set
by the disk.

Our procedure follows that of \citet[][PY]{2014ApJ...786...21P}, with
a few exceptions detailed below, such as allowing for gradients in
composition and more complex opacities. The basic idea is that as the
envelope cools, it contracts and accretes more gas from the disk. But
on timescales shorter than the accretion (i.e., cooling or
Kelvin-Helmholtz) time, the atmosphere is practically
hydrostatic. Thus we begin by constructing a series of ``hydrostatic
snapshots'' (\S\ref{ssec:equations}):
1D atmospheric models in strict hydrostatic equilibrium,
each having a unique gas-to-core mass ratio GCR $\equiv M_{\rm
  gas}/M_{\rm core}$.
We then string these snapshots together
in time (in order of increasing GCR) by calculating the rate at
which the planet cools from one snapshot to the next
(\S\ref{ssec:time_sequence}).

\subsection{Hydrostatic Snapshots}
\label{ssec:equations}

Each hydrostatic snapshot is constructed by solving the standard equations
of stellar structure:

\begin{equation}
\frac{dM}{dr} = 4\pi r^2\rho
\label{eq:mass_conserv}
\end{equation}

\begin{equation}
\frac{dP}{dr} = -\frac{GM(<r)}{r^2}\rho
\label{eq:HSE}
\end{equation}

\begin{equation}
\frac{dT}{dr} = \frac{T}{P} \frac{dP}{dr} \nabla
\label{eq:nabla}
\end{equation}

\noindent where $r$ is radius, $\rho$ is density, $P$ is pressure, $G$
is the gravitational constant, $M(<r)$ is the mass enclosed within
$r$, and $T$ is temperature.  The dimensionless temperature gradient
$\nabla \equiv d\ln T / d\ln P$ equals either 

\begin{equation}
\label{eq:grad_rad}
\nabla_{\rm rad} = \frac{3\kappa P}{64\pi G M \sigma T^4} L
\end{equation}

\noindent where energy transport is by radiative diffusion, or

\begin{equation}
\label{eq:grad_ad} 
\nabla_{\rm ad} = -\left.\frac{\partial\log S}{\partial\log
  P}\right\vert_T\left(\left.\frac{\partial\log S}{\partial\log T}\right\vert_P\right)^{-1}
\end{equation}

\noindent where transport is by convection. Here $L$ is luminosity,
$S$ is the specific entropy (\S\ref{sssec:EOS}),
$\kappa$ is opacity (\S\ref{sssec:opacity}), and $\sigma$ is the
Stefan-Boltzmann constant. Convection prevails according to the
Ledoux criterion:
\begin{equation}
\nabla_{\rm rad} - \nabla_{\mu} > \nabla_{\rm ad} \,\,\,\,\,\,\,\,\,\, ({\rm unstable \,\,to \,\,convection})
\label{eq:ledoux}
\end{equation}
where $\mu$ is the mean molecular weight and $\nabla_{\mu}
\equiv d\ln\mu/d\ln P$ accounts for compositional gradients. 
 Assuming the heavy elements are homogeneously distributed,
we find that $\nabla_{\mu}$ is negative and so acts to drive convection;
however, the effect is
negligible. (In models of Saturn and Jupiter,
$\nabla_{\mu}$ can be positive because of imperfect mixing of 
solids, immiscibility of helium, and core erosion; see \citealt{leconte12, leconte13}.
These effects manifest at pressures
and densities that prevail only at the very bottoms of super-Earth atmospheres.)

A major simplification in our procedure is to assume that $L$ is spatially
constant (e.g., PY). The assumption is valid in
radiative zones if their thermal relaxation times are shorter than 
thermal times in the rest of the atmosphere. Then before the planet
can cool as a whole (i.e., before one snapshot transitions to another),
its radiative zones relax to a thermal steady state in
which energy flows outward at a constant rate. We will check whether
this is the case in \S\ref{ssec:fiducial}. By comparison
in convective zones, the assumption of
constant $L$ is moot, because there the density and temperature
structures follow an adiabat, $P \propto \rho^{\gamma}$ where $\gamma
= (1-\nabla_{\rm ad})^{-1}$, independent of $L$.

In computing a snapshot for a desired GCR, the spatially constant $L$
is an eigenvalue found by iteration.  We guess $L$, integrate
(\ref{eq:mass_conserv})--(\ref{eq:nabla}) inward from a set of outer
boundary conditions until we reach the core radius $R_{\rm core}$, and
compute the resultant GCR. The integration is performed using Python's
\texttt{odeint} package. If the GCR does not match the one desired,
then we repeat the integration with a new $L$. We iterate on $L$ until
the desired GCR is reached.

As a simplifying measure, we neglect whatever intrinsic luminosity 
and heat capacity 
the rocky core may have. The validity of this approximation will be assessed
when we present the results for our fiducial model in \S\ref{ssec:fiducial}.

\subsubsection{Boundary conditions}
\label{sssec:BC}

The base of the atmosphere is located at the surface of the solid
core, whose bulk density is fixed at $\rho_{\rm core} = 7$ g/cm$^3$.
For a fiducial core mass of $M_{\rm core} = 5M_{\oplus}$,
we have $R_{\rm core} = 1.6 R_\oplus$.

The outer radius $R_{\rm out}$ is set either to the Hill radius
\begin{eqnarray}
R_{\rm H} &=& \left[\frac{(1+{\rm GCR})M_{\rm core}}{3M_{\odot}}\right]^{1/3}a \nonumber \\
    &\simeq& 40 R_{\oplus} \left[\frac{(1+{\rm GCR})M_{\rm core}}{5M_{\oplus}}\right]^{1/3}\left(\frac{a}{0.1\,\rm{AU}}\right),
\label{eq:Rhill}
\end{eqnarray}
or the Bondi radius
\begin{eqnarray}
R_{\rm B} &=& \frac{G(1+{\rm GCR})M_{\rm core}}{c_{\rm s}^2} \nonumber \\
        &\simeq& 90 R_{\oplus} \left[\frac{(1+{\rm GCR})M_{\rm core}}{5M_{\oplus}}\right] \left( \frac{\mu}{2.37} \right)
\left(\frac{1000~{\rm K}}{T}\right) \,,
\label{eq:Rb}
\end{eqnarray}
whichever is smaller. Here $c_{\rm s}$, $T$, and $\mu$
are the sound speed, temperature, and mean molecular weight of disk
gas at stellocentric distance $a$. We fix the host stellar mass to be $1 M_\odot$.
The temperature above which $R_{\rm B} \leq R_{\rm H}$ is 
\begin{equation}
T_{\rm HB} \simeq 2200~{\rm K} \left[\frac{(1+{\rm GCR})M_{\rm core}}{5M_{\oplus}}\right]^{2/3} \left( \frac{\mu}{2.37} \right)
\left(\frac{0.1\,{\rm AU}}{a}\right) \,.
\label{eq:Tbondi}
\end{equation}
Our fiducial input parameters $T (R_{\rm out}) = 10^3$ K and gas
density $\rho (R_{\rm out}) = 6 \times 10^{-6}$ g/cm$^3$
are drawn from the minimum-mass extrasolar nebula (MMEN) of
\citet{2013MNRAS.431.3444C}:
\begin{equation}
\rho_{\rm MMEN} = 6\times 10^{-6} \left( \frac{a}{0.1\,{\rm AU}} \right)^{-2.9}~\rhocgs
\end{equation}
\begin{equation}
T_{\rm MMEN} = 1000 \left( \frac{a}{0.1 \,{\rm AU}} \right)^{-3/7} \K
\end{equation}
where the power-law index on temperature is taken from \citet{chiang97}.
Other disk models (e.g.,
\citealt{2006ApJ...648..666R}; \citealt{2012ApJ...751..158H}) yield
similar outer boundary conditions. We will perform a parameter study
over $M_{\rm core}$, $T (R_{\rm out})$, $\rho (R_{\rm out})$, and
$\kappa (R_{\rm out})$ in \S\ref{ssec:dep_on_bc}. For a given
model, nebular parameters are fixed in time; however, this simplification
will not prevent us from making rough connections between our models
and dissipating (time-variable) disks in \S\ref{sec:disc}.

If $R_{\rm out}$ exceeds the disk scale height $H = c_{\rm s}
a^{3/2}/\sqrt{GM_\odot}$, our assumption of spherical symmetry breaks
down. For our
fiducial parameters at $a \sim 0.1$ AU, we have $H \simeq 50
R_\oplus$ which is comparable to $R_{\rm out} = \min (R_{\rm H},
R_{\rm B}) \simeq 40 R_\oplus$. Thus the error we accrue by ignoring the
disk's vertical density gradient seems at most on the order of unity
--- but this assessment is subject to errors in our prescription
itself for $R_{\rm out}$, which ignores how the true radius inside of
which material is bound to the planet may differ from $R_{\rm H}$ or
$R_{\rm B}$.
We will explore the sensitivity of our results to $R_{\rm
  out}$ in \S\ref{sssec:Rout}.

\subsubsection{Equation of state}
\label{sssec:EOS}

We compute our own ideal-gas equation of state (EOS) for a mixture of
hydrogen, helium, and metals. 
For given temperature $T$ and pressure
$P$, the EOS yields density $\rho$, adiabatic temperature gradient
$\nabla_{\rm ad}$, and internal energy $U$. The internal energy is
used only to connect our hydrostatic snapshots in time
(\S\ref{ssec:time_sequence}), and does not enter into the construction
of an individual snapshot.
Our model has an advantage over
the commonly used \citet{1995ApJS...99..713S} EOS
as we account for metals (albeit crudely), 
but has the
disadvantage that we do not consider quantum effects and
intermolecular interactions. Fortunately, these omissions
are minor for super-Earth atmospheres (\S\ref{sssec:checks}).
For our fiducial model, 
values of $\nabla_{\rm ad}$ calculated from our ideal-gas EOS agree with those of
\citet{1995ApJS...99..713S} to within $10\%$, with similar levels of agreement 
for $\rho$ and $U$ except in a few locations where discrepancies approach factors of 2.
We have checked that
mechanical and thermal stability are satisfied over the parameter
space relevant to our study.

The density of our mixture is given by
\begin{equation}
\rho = \frac{P m_{\rm H}}{kT \left(X/\mu_{\rm H} + Y/\mu_{\rm He} + Z/\mu_{\rm Z} \right)}
\label{eq:rho}
\end{equation}
where $k$ is the Boltzmann constant and $m_{\rm H}$ is the mass of the
hydrogen atom. For (our fiducial) solar composition,
we use the mass fractions $X=0.7$, $Y=0.28$, $Z=0.02$.
We also explore subsolar
($X=0.713$, $Y=0.285$, $Z=0.002$) and supersolar ($X=0.57$, $Y=0.23$,
$Z=0.2$) compositions.
Both helium and metals are assumed to remain
atomic throughout the atmosphere;\footnote{
In reality, metals can take the form of molecules.
At the temperatures
$T \gtrsim 2000$ K characterizing close-in super-Earths,
CO is the dominant molecule \citep{2011MNRAS.416.1419H}.
We have verified by direct calculation that for $Z \lesssim 0.2$,
our results hardly change whether all of the cosmic abundance
of C is atomic or whether it is locked up in CO.
Dissociation of CO occurs only at the very bottom of our
atmospheres (at the core-envelope boundary),
and for $Z \lesssim 0.2$ there is not enough CO 
to significantly alter $\nabla_{\rm ad}$.
At lower temperatures $T \simeq 100$--1000 K and $Z \gtrsim 0.5$,
molecular metals have greater impact: the increased molecular
weight and the presence of H$_2$O---with its many degrees of freedom
and its potential for reacting chemically---render envelopes
more susceptible to runaway \citep{2011MNRAS.416.1419H}.\label{foot:molecule}}
accordingly,
we fix $\mu_{\rm He} = 4$ and $\mu_{\rm Z} = 16.95$
(representing an average over the relative metal abundances
tabulated by \citealt{GN93}). We have checked
{\it a posteriori} that our neglect of helium ionization
is safe because its effects are felt only at the very bottom
of our atmosphere, near the core surface.
For hydrogen, we distinguish
between its ionized, atomic, and molecular forms:
\begin{equation}
\mu_{\rm H} = \frac{X}{2X_{\rm HII} + X_{\rm HI} + X_{\rm H_2}/2}
\label{eq:mu}
\end{equation}
where we have adopted the convention that the mass fractions $X_{\rm HII} + X_{\rm HI} + X_{\rm H_2} = X$.

We compute the mass fractions $\{ X_{\rm HII}, X_{\rm HI}, X_{\rm H_2} \}$
via the corresponding number fractions.
In thermodynamic equilibrium, the atomic number fraction $x_{\rm HI}
\equiv n_{\rm HI}/n_{\rm tot}$ 
is given through Saha-like considerations by
\begin{equation}
\frac{x_{\rm HI}^2}{1-x_{\rm HI}} = \frac{Z_{\rm tr,HI}^2}{Z_{\rm
    tr,H_2}}\frac{Z_{\rm int,HI}^2}{Z_{\rm int,H_2}}
e^{-E_{\rm b}/kT}
\label{eq:dissoc}
\end{equation}
where $E_{\rm b}=4.5167
\,{\rm eV}$ is the binding energy of $\rm{H_2}$, 
$Z_{\rm tr,HI}=(2\pi m_{\rm H}kT/h^2)^{3/2}/n_{\rm HI}$ and
$Z_{\rm tr,H_2}=(4\pi m_{\rm H}kT/h^2)^{3/2}/n_{\rm H_2}$ are the 
translational partition functions for atomic and molecular hydrogen,
respectively, and $Z_{\rm int,HI}$
and $Z_{\rm int,H_2}$ are internal partition functions defined below.
Only in equation (\ref{eq:dissoc}) do we approximate the total number density
$n_{\rm tot} = n_{\rm HI} + n_{\rm H_2}$; for purposes of evaluating
$x_{\rm HI}$, we assume that $n_{\rm HII}/n_{\rm tot}$ is small and
neglect other trace elements. However, $x_{\rm HII} \equiv n_{\rm HII}/n_{\rm HI}$ is not necessarily negligible, and derives from the Saha equation:
\begin{equation}
\frac{x_{\rm HII}^2}{1-x_{\rm HII}} = \frac{Z_{\rm tr,p}Z_{\rm
    tr,e}}{Z_{\rm tr,HI}}\frac{4}{Z_{\rm
    int,HI}}e^{-13.6\, {\rm eV}/kT}
\label{eq:saha}
\end{equation}
where $Z_{\rm tr,p}=(2\pi m_{\rm p}kT/h^2)^{3/2}/n_{\rm HII}$ and
$Z_{\rm tr,e}=(2\pi m_{\rm e}kT/h^2)^{3/2}/n_{\rm HII}$ are the
translational partition functions for free protons and electrons,
respectively. The factor of 4 accounts for electron and proton spin
degeneracies.

Only electronic states contribute to the internal partition
function for atomic hydrogen:
\begin{equation}
Z_{\rm int,HI} = 4\sum^{n_{\rm max}}_{n=1}n^2 e^{-13.6 \,{\rm eV}(1-1/n^2)/kT} \,.
\label{eq:Za}
\end{equation}
We choose the maximum quantum number
$n_{\rm max}$ such that the effective radius of the outermost electron shell
$n_{\rm max}^2 a_0$ equals half the local mean particle spacing
$n_{\rm tot}^{-1/3}$, where $a_0$ is the Bohr radius.
The internal partition function for molecular hydrogen is
\begin{equation}
Z_{\rm int,H_2} = Z_{\rm elec,H_2} Z_{\rm vib,H_2} Z_{\rm
  rot,H_2}
\end{equation}
where
\begin{equation}
Z_{\rm elec,H_2} \simeq 1 + e^{-E_{\rm b}/kT},
\label{eq:Zmelec}
\end{equation}
\begin{equation}
Z_{\rm vib,H_2} = \sum^{\infty}_{n=0}e^{- 0.546 \, {\rm eV} (n+1/2) / kT},
\label{eq:Zmvib}
\end{equation}
and
\begin{equation}
\begin{split}
Z_{\rm rot,H_2} &= Z_{\rm para} + 3Z_{\rm ortho} \\
           &=\sum_{\rm{even}~\rm{j}}(2j+1)e^{-j(j+1)\hbar^2/2IkT} \\
					&\quad+3\sum_{\rm{odd}~\rm{j}}(2j+1)e^{-j(j+1)\hbar^2/2IkT}
\end{split}
\label{eq:Zmrot}
\end{equation}
with $\hbar = h/2\pi$ and $I=4.57\times 10^{-41}\g\cm^2$, 
 and where the prefactors 1 and 3 in equation (\ref{eq:Zmrot})
denote the nuclear spin degeneracies.
Note that we do not assume a fixed
ortho-to-para ratio for molecular hydrogen,
but let the ratio vary with temperature $T$ in thermal
equilibrium.

To compute $\nabla_{\rm ad}$ (equation \ref{eq:grad_ad}),
we tabulate the specific entropy $S$ on a logarithmically evenly spaced grid of temperature
and pressure with $d\log T=0.02$ and $d\log P=0.05$. To calculate
derivatives, we take local cubic splines of $S$ on small square patches
that are 7 grid spacings on each side. The entropy per mass is evaluated as
\begin{equation}
\begin{split}
S = &(X_{\rm e} S_{\rm e} + X_{\rm p}S_{\rm p} + X_{\rm HI}S_{\rm
  HI} + X_{\rm H_2}S_{\rm H_2})\\
	&+YS_{\rm He}+ZS_{\rm Z}+S_{\rm mix}
\end{split}
\label{eq:S}
\end{equation}
where $S_{\rm mix}$ is the entropy of mixing (e.g., \citealt{1995ApJS...99..713S}):
\begin{equation}
S_{\rm mix}/k = {\cal N}\log{\cal N} -\sum_i \frac{X_i}{m_i}\log\left(\frac{X_i}{m_i}\right)
\label{eq:Smix}
\end{equation}
with
\begin{equation}
{\cal N} = \frac{X}{\mu_{\rm H}m_{\rm H}}+\frac{Y}{\mu_{\rm He}m_{\rm
    H}}+\frac{Z}{\mu_{\rm Z}m_{\rm H}} \,.
\label{eq:scriptN}
\end{equation}
The index $i$ iterates over free electrons, free protons, atomic
hydrogen, molecular hydrogen, helium, and metals. For species $i$, the
particle mass is $m_i$ and the mass fraction is $X_i$ (i.e.,
$\sum_{i=1}^6 X_i = 1$; note that $X_{\rm He} \equiv Y$, $X_{\rm Z} \equiv Z$,
and $X_{\rm p} + X_{\rm e} \equiv X_{\rm HII}$).
The entropy of an individual species is
calculated from its Helmholtz free energy $F$. For example, for
atomic hydrogen,
\begin{equation}
S_{\rm HI} = -\left.\frac{\partial F_{\rm HI}}{\partial T}\right\vert_{\rho,\mu}
\label{eq:S_H}
\end{equation}
where
\begin{equation}
F_{\rm HI} = -\frac{kT}{m_{\rm H}}\log (Z_{\rm tr,HI}Z_{\rm int,HI}) \,.
\label{eq:F}
\end{equation}
For molecular hydrogen, we account for electronic, vibrational, and
rotational partition functions; for protons and electrons, we account
for their spin degeneracies and their translational partition
functions; and for helium and metals, we account only for their
translational partition functions.

Finally, the total internal energy $U = (X_{\rm e} U_{\rm e} + X_{\rm
  p}U_{\rm p}+X_{\rm HI}U_{\rm HI}+X_{\rm H_2}U_{\rm H_2}) + YU_{\rm
  He} + ZU_{\rm Z}$ where the internal energy of each species can be
derived from $F$ and $S$: e.g., $U_{\rm HI} = F_{\rm HI} + TS_{\rm
  HI}$. Note that the total free energy from all hydrogenic species is
not merely the sum of the individual free energies $F_{\rm H_2} +
F_{\rm HI} + F_{\rm HII}$, because $S$ does not add linearly.

\subsubsection{Opacity}
\label{sssec:opacity}

Piso \& Youdin (2014) have stressed the importance of opacity in
regulating the accretion (read: cooling) history of gas giant cores,
and our work will prove no exception. For close-in super-Earth
atmospheres, we need opacities $\kappa$ over the following ranges of
densities and temperatures: $-6 < \log \rho \,({\rm g} / {\rm cm}^3) <
-1$ and $2.7 < \log T\, ({\rm K}) < 4.5$.  To this end, we utilize the
opacity tables of \citet{2005ApJ...623..585F}, which partially span
our desired ranges, and interpolate/extrapolate where necessary.

\begin{table*}[!hbt]
\caption{Opacity Fit Parameters ($\kappa = \kappa_i \rho^{\alpha_i} T^{\beta_i}$; all quantities in cgs units)}
\label{tab:opacity}
\begin{center}
\begin{tabular}{ c c c c c c c c c c c }
\hline \hline
$Z$ & Dust & $\log\kappa_{b}$ & $\alpha_{b}$ & $\beta_{b}$ &
$\log\kappa_{x1}$ & $\alpha_{x1}$ & $\beta_{x1}$ &
$\log\kappa_{x2}$ & $\alpha_{x2}$ & $\beta_{x2}$ \\
\hline
0.02  & yes & -22 & 0.46 & 6.7 & -25 & 0.53 & 7.5 & -13 &
              0.46 & 4.5 \\
      & no  & -25 & 0.50 & 7.5 & -25 & 0.53 & 7.5 & -13 &
              0.46 & 4.5 \\
0.002 & yes & -27 & 0.49 & 7.8 & -28 & 0.60 & 8.2 & -14 &
              0.48 & 4.8 \\
      & no  & -28 & 0.55 & 8.1 & -28 & 0.60 & 8.2 & -14 &
              0.48 & 4.8 \\
0.2   & yes & -23 & 0.48 & 7.1 & -18 & 0.48 & 5.9 & -6.1 &
              0.40 & 2.8 \\
      & no  & -22 & 0.48 & 7.0 & -19 & 0.35 & 6.0 & -6.1 &
              0.40 & 2.8 \\
\hline
\end{tabular}
\end{center}
\end{table*}

We experiment with a total of 6 opacity laws: 3 metallicities (solar
$Z_{\odot} = 0.02$, subsolar $0.1 Z_\odot$, and supersolar
$10 Z_\odot$,\footnote{The abundances used for our supersolar opacity 
model are slightly discrepant from those of our EOS model (\S\ref{sssec:EOS}); 
the former uses $\{X=0.53, \,Y=0.27, \,Z=0.2\}$, whereas the latter uses
$\{X=0.57, \,Y=0.23, \,Z=0.2\}$. The opacity and EOS models also differ 
generically in that the opacity model (drawn directly from 
\citealt{2005ApJ...623..585F}) uses its own equation of state
based on the \texttt{PHOENIX} stellar atmospheres code, which includes 
molecular metals. These differences are not significant; we have verified that changing the hydrogen/helium 
abundances in our supersolar EOS by several percent to reconcile them 
with our supersolar opacity model changes our computed times to runaway 
gas accretion by $\lesssim 15\%$; see also footnote \ref{foot:molecule}.}
where elemental abundances are scaled to those in
\citealt{GN93}) $\times$ 2 dust models (``dusty'' which assumes the
ISM-like grain size distribution of \citealt{2005ApJ...623..585F}; and
``dust-free'' in which
metals never take the form of dust and are instead in the gas phase
at their full assumed abundances).
Where the Ferguson et al.~(2005) tables are incomplete, we interpolate
or extrapolate using power laws. Our look-up table
is constructed as follows (with $T$ in K and $\rho$ in g/cm$^3$):
\begin{enumerate}
\item $\log T \geq 3.65$ and $\log\rho \leq -3$: tabulated in \citet{2005ApJ...623..585F}, with supplemental data calculated by J.~Ferguson (2013, personal communication)
\item $3.6 < \log T < \log T_{\rm blend}$: $\kappa = \kappa_{b}~\rho^{\alpha_{b}}~T^{\beta_{b}}$ (interpolation)
\item $2.7 \leq \log T \leq 3.6$ and $\log\rho \geq -6$: tabulated in \citet{2005ApJ...623..585F}, with supplemental data calculated by J.~Ferguson (2013, personal communication)
\item $\log T < 2.7$ and $\log\rho < -6$: $\kappa = \kappa(\log T = 2.7, \log\rho = -6)$ (constant extrapolation)
\item $T \leq T_{\rm hi}$ and $\log\rho > -3$: $\kappa = \kappa_{x1}~\rho^{\alpha_{x1}}~T^{\beta_{x1}}$ (extrapolation)
\item $T > T_{\rm hi}$ and $\log\rho > -3$: $\kappa = \kappa_{x2}~\rho^{\alpha_{x2}}~T^{\beta_{x2}}$ (extrapolation)
\end{enumerate}
where $\log T_{\rm blend} = 3.75$ for supersolar metallicity, 3.7 for all dust-free models, 
and 3.65 otherwise;
and $\log T_{\rm hi} = 4.2$ for supersolar metallicity and 3.9 otherwise.
Table 
\ref{tab:opacity} lists our fit parameters. We have verified that our use of a constant extrapolation at low $\rho$ and low $T$ (item 4 above) is acceptable. 
For $\log T < 2.7$, we expect the opacity to be dominated by dust and 
independent of $\rho$, assuming a constant dust-to-gas ratio. 
We have confirmed that $\kappa$ indeed hardly varies with $\rho$ at these low 
temperatures (J.~Ferguson 2014, personal communication).
We have also checked that our results are insensitive to different extrapolated 
temperature scalings for $\log T < 2.7$ (as steep as $\kappa \propto T^2$).

\begin{figure}
\centering
\includegraphics[width=0.5\textwidth]{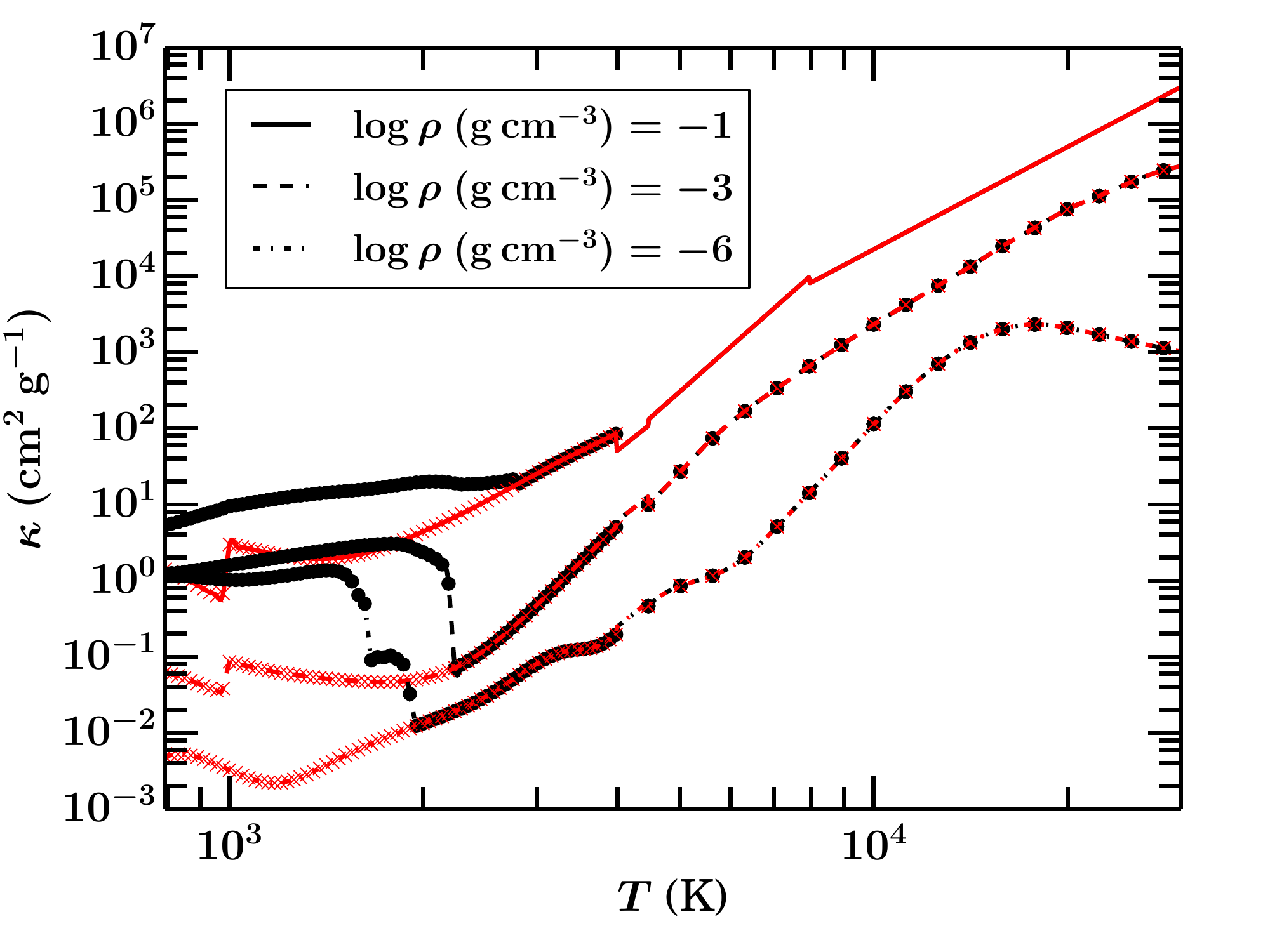}
\caption{\label{fig:kappa}Tabulated and extrapolated dusty (black) and
  dust-free (red) solar metallicity opacities vs.~temperature at
  several densities. Symbols
  correspond to tabulated values while continuous curves correspond to
  extrapolated values.}
\end{figure}

Figure \ref{fig:kappa} illustrates how our solar metallicity +
dusty (and dust-free) $\kappa$ varies with $T$ for representative values of
$\rho$.
Dust, when present, dominates at $T \lesssim 1700$ K.
For $1700 \K \lesssim
T \lesssim 2300 \K$, dust sublimates, one grain species at a time according to
its condensation temperature, leaving behind gas molecules as the
primary source of opacity. Molecular opacity sets a floor on $\kappa$
of $\sim$$10^{-2}$ cm$^2$/g, two orders of magnitude below the maximum
dust opacity.
We emphasize that our opacity model
keeps track of how the gas phase abundances of the
refractory elements vary with temperature according to the 
sublimation fronts of various grain species.

For $2500 \K \lesssim T \lesssim 18000 \K$, H$^-$ ions
provide most of the opacity 
(with contributions from H$_2$-H$_2$ and H$_2$-He collision-induced absorption;
cf.~\citealt{guillot94}).
The H$^-$ opacity rises steeply with $T$,
reflecting the growing abundance of H$^-$ with increasing atomic
fraction $x_{\rm HI}$ and increasing numbers of free electrons from
thermally ionized species.

\subsection{Connecting Snapshots in Time}
\label{ssec:time_sequence}

When the gas envelope lacks an internal power source (from, e.g.,
fusion or planetesimal accretion), the time $\Delta t$ between two
successive hydrostatic snapshots is simply the time it takes to cool
from one to the other, in order of increasing GCR. This cooling time
is modified slightly by changes to the energy budget from gas
accretion and envelope contraction. From PY (see their Appendix A for
a derivation), we have
\begin{equation}
\Delta t = \frac{-\Delta E + \langle e_M \rangle \Delta M - \langle P
  \rangle \Delta
  V_{\langle M \rangle}}{\langle L \rangle}
\label{eq:dt}
\end{equation}
where $\langle Q \rangle$ denotes the average of quantity $Q$
in two adjacent snapshots, and $\Delta Q$ denotes the
difference between snapshots. The luminosity $L$ is the eigenvalue
satisfying the equations of stellar structure, found by iteration
as described in \S\ref{ssec:equations}.
From left to right, the terms in the numerator of (\ref{eq:dt})
account for (a) the change in total (gravitational plus internal) energy 
\begin{equation}
E = -\int \frac{GM(<r)}{r}dM + \int U dM
\label{eq:E}
\end{equation}
integrated over the innermost convective zone ($\Delta E < 0$; note
that $M(<r)$ takes the core mass into account);
(b) the energy accrued by accreting gas ($\Delta M > 0$) with
specific energy
\begin{equation}
e_M = -\left.\frac{GM}{r}\right|_{R_{\rm RCB}} + \left.
U
\right|_{R_{\rm RCB}} 
\label{eq:eM}
\end{equation}
where $R_{\rm RCB}$ is the radius of the innermost
radiative-convective boundary (RCB); and finally, (c) the work done on
the planet by the contracting envelope, with $\Delta V_{\langle M
  \rangle} < 0$ equal to the change in the volume enclosing the
average of the innermost convective masses of the two snapshots,
and $\langle P \rangle$ equal to the averaged
pressure at the surface of this volume. We emphasize that we evaluate all
three terms at the boundary of the innermost convective zone, as this
seems the most natural choice given our assumption that all the
luminosity is generated inside \citep{2014ApJ...786...21P}.

Because the procedure above only yields changes in time $\Delta t$
between snapshots, we still need to specify a time $t_0$ for the first
snapshot.  In practice, the first snapshot is that for which the
atmosphere is nearly completely convective, since $L$ cannot be found
for fully convective atmospheres.\footnote{In reality, a thin
  radiative layer should always cap the planet's atmosphere and
  regulate the loss of energy from the planet to the disk. Resolving
  this outer boundary layer---which may also advect energy to the disk---is
  a forefront problem. Its
  solution would enable us to probe still earlier times and smaller
  GCRs than we can reach in this paper. \label{foot:outer}}
For this first snapshot we assign $t_0 := |E|/L$.
Whatever formal error is accrued
in making this assignment is small
insofar as $t_0$ is much less than the times
to which we ultimately integrate (e.g., the time of runaway gas accretion).

\section{RESULTS}
\label{sec:results}

The time evolution of our fiducial model of a growing super-Earth
atmosphere is described in \S\ref{ssec:fiducial},
together with an explanation of our findings to order-of-magnitude
accuracy (\S\ref{sssec:oom}) and a review of the {\it a posteriori}
checks we performed (\S\ref{sssec:checks}).  How runaway gas accretion
is promoted or inhibited by varying nebular conditions and core masses
is surveyed in \S\ref{ssec:dep_on_bc}.

\begin{figure}[!htb]
\centering
\includegraphics[width=0.5\textwidth]{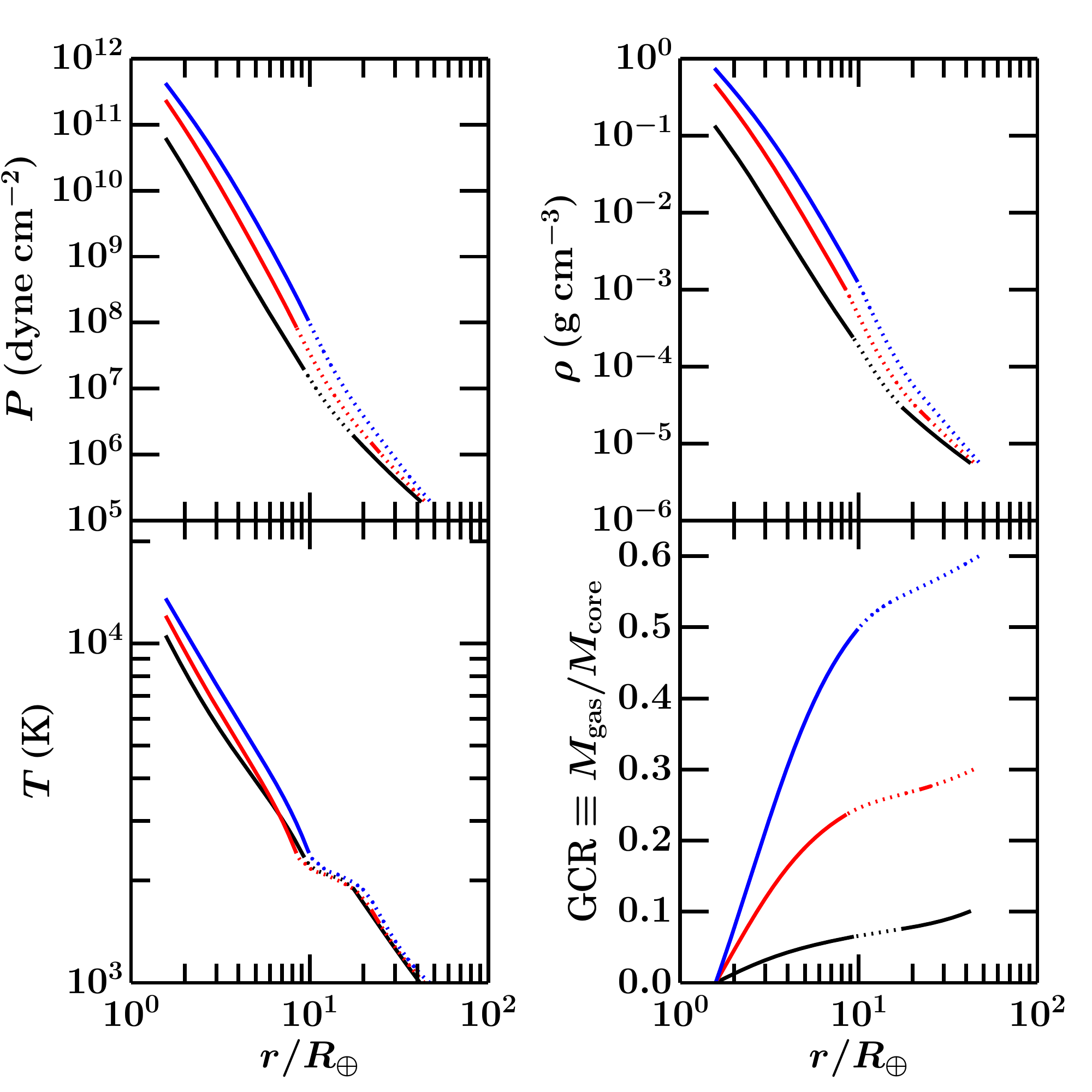}
\caption{\label{fig:atm_prof} Atmospheric profiles for GCR=0.1
  (black), GCR=0.3 (red), and GCR=0.6 (blue) in our fiducial
  model. Dotted lines trace radiative zones while solid lines trace
  convective zones.}
\end{figure}

\subsection{Fiducial Model}
\label{ssec:fiducial}

Our fiducial model is a 5$M_{\oplus}$, 1.6$R_{\oplus}$ solid core
located at $a=0.1$ AU in the minimum-mass extrasolar nebula (MMEN) 
with $\rho = \rho_{\rm MMEN} = 6\times 10^{-6}~(a/0.1\,{\rm AU})^{-2.9}~\rhocgs$ and $T = T_{\rm MMEN} = 1000~(a/0.1 \,{\rm AU})^{-3/7}$ K.
We assume the disk to be dusty, with solar metallicity and an ISM-like grain size distribution.
In Figure \ref{fig:atm_prof}, we show how various atmospheric properties vary
with depth for different envelope masses. 
Most of the atmosphere is in
the innermost convective zone---at least 75\% by mass for GCR $\geq$ 0.2.
The outermost
layer is always cool enough for dust to survive and dominate the
opacity; at small GCRs (early times), this outer dusty layer is
convective; at higher GCRs, it becomes marginally radiative (the
temperature profile remains nearly adiabatic). Sandwiched between
this outermost layer and the innermost convective zone is a radiative
layer so hot that dust sublimates and where the opacity is at a global
minimum. Temperature profiles in this radiative zone are shallower,
and consequently pressure and density profiles are steeper, than in
convective zones.

\begin{figure}[!htb]
\centering
\includegraphics[width=0.45\textwidth]{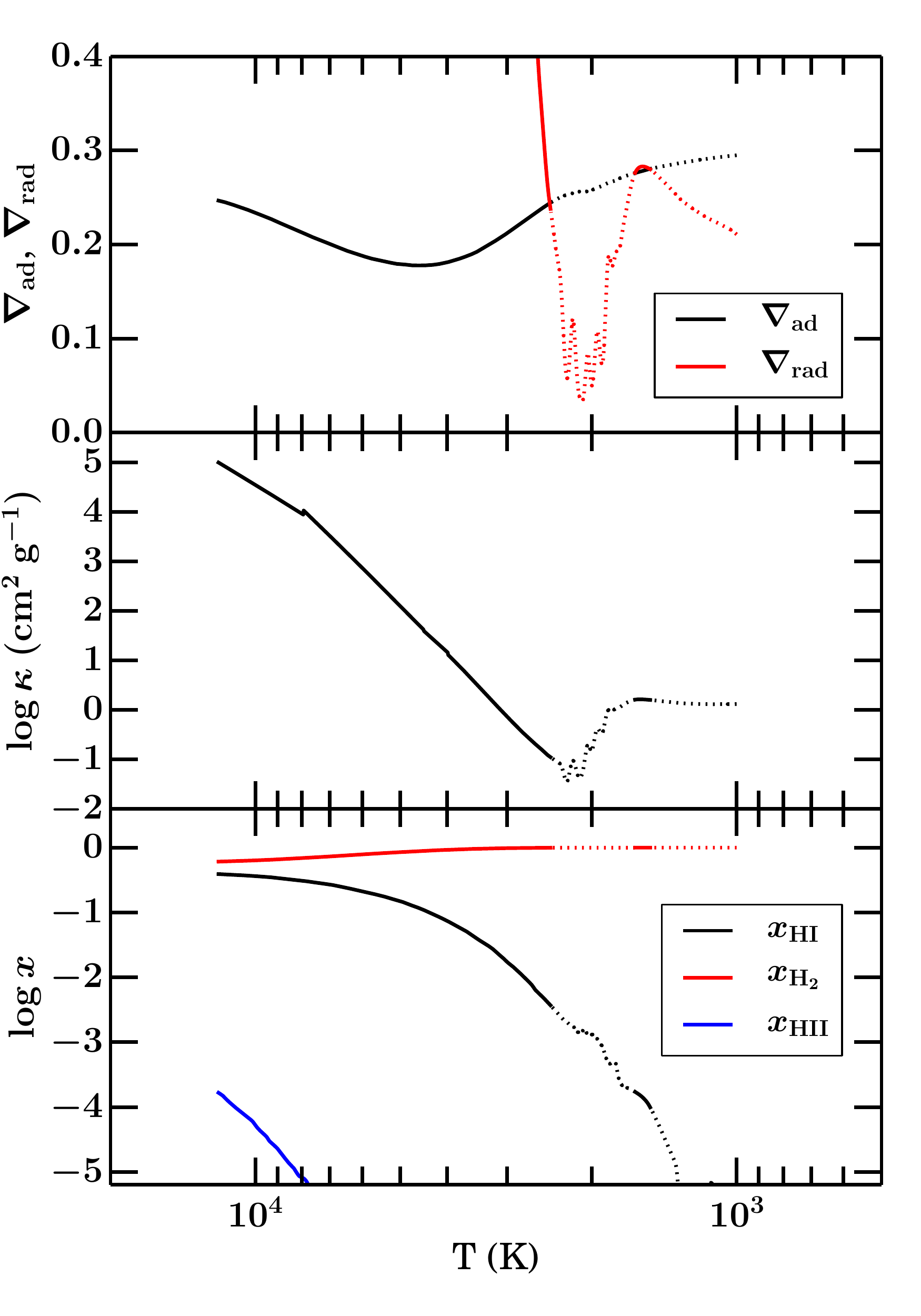}
\caption{\label{fig:gradad} Adiabatic (black) and
  radiative (red) temperature gradients
(top), opacity (middle), and number fractions of different hydrogen species
(bottom) vs.~temperature for GCR=0.4 in our fiducial model. Dotted lines
trace radiative zones while solid lines trace convective
zones. The location of the innermost 
RCB is determined by the ${\rm H_2}$ dissociation front. 
The oscillations in $\nabla_{\rm rad}$ and $\kappa$ are due to different dust species 
evaporating at different temperatures.}
\end{figure}

\begin{figure}[!htb]
\centering
\includegraphics[width=0.5\textwidth]{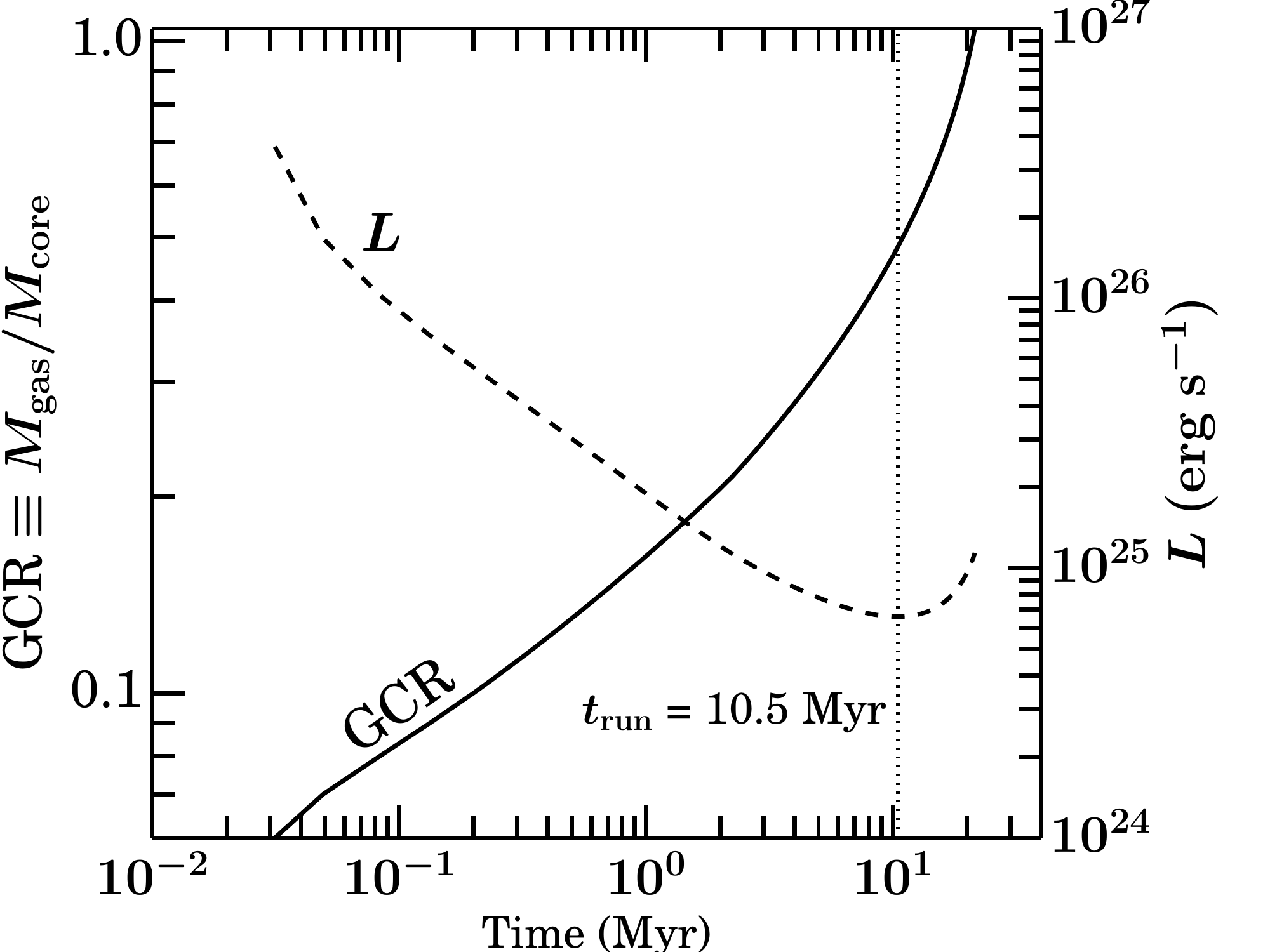}
\caption{\label{fig:L_eta}Time evolution of GCR (solid curve) and $L$
  (dashed curve) for our fiducial model of a 5 $M_\oplus$ core at 0.1 AU. 
  The dotted vertical
  line denotes $t_{\rm run}\simeq$ 10.5~Myr, defined as the time when
  $L$ reaches its minimum. The GCR starts to rise superlinearly after
  minimum $L$, signalling
  runaway accretion. The initial time is taken as $t_0 = |E|/L
  \simeq 0.03$ Myr at the lowest GCR of 0.06, below which the
  atmosphere becomes completely convective and the evolution cannot be followed.
}
\end{figure}

\begin{figure}[!hbt]
\centering 
\includegraphics[width=0.5\textwidth]{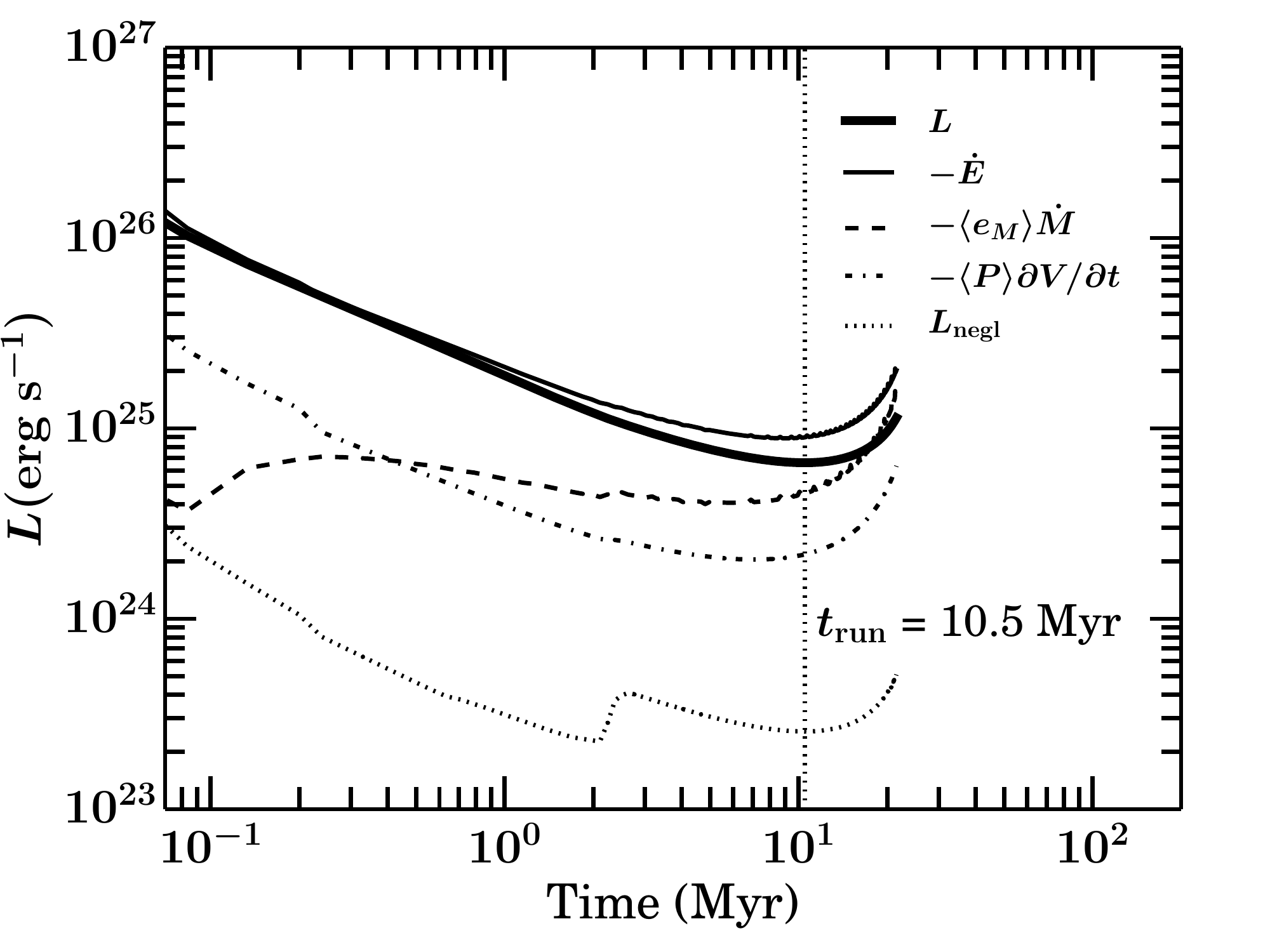}
\caption{\label{fig:dt_contr}Contributions to the atmospheric power budget (equation
  \ref{eq:dt}) vs.~time for our fiducial model. Also plotted is
  the luminosity neglected in radiative zones (dotted curve). The sudden increase in
  $L_{\rm negl}$ at $\sim$2 Myr is due to the emergence of an outer
  radiative zone. Changes in total energy ($\dot{E}$) provide almost
  all the luminosity $L$ at least until $t_{\rm run}$ (dotted vertical
  line), when the energy
  input from surface accretion $\langle e_M \rangle \dot{M}$ becomes significant.}
\end{figure}

Figure \ref{fig:atm_prof} shows that the temperature at the innermost
RCB (the RCB from hereon) stays at $\sim$2500~K at all times. This is
the temperature at which ${\rm H_2}$ begins to dissociate (equation
\ref{eq:dissoc}). As Figure \ref{fig:gradad} explains, the transition
from a radiative to a convective zone is caused by the decrease in
$\nabla_{\rm ad}$ and the steep increase in 
$\nabla_{\rm rad}$,
both brought about by
${\rm H_2}$ dissociation. At the dissociation front, the gas temperature tends
to stay fixed as energy is used to break up molecules rather than to
increase thermal motions. This near-isothermal behavior drives
$\nabla_{\rm ad}$ downward, facilitating the onset of convection. The
creation of H atoms also allows the formation of ${\rm H^-}$ ions, the
dominant source of opacity for $T\gtrsim 2500$ K. The surge of opacity
from ${\rm H^-}$, together with the near-constant temperature profile,
increases $\nabla_{\rm rad}$ and causes radiation to give way to convection
as the dominant transport mechanism. 

The evolution of luminosity is displayed in Figure
\ref{fig:L_eta}. Initially $L$ falls.
The drop in radiative luminosity occurs as density and pressure---and
therefore optical depth---rise at the RCB
(compare, e.g., the GCR=0.1 and GCR=0.3 profiles for $P$ and $\rho$
in Figure \ref{fig:atm_prof}). 
As revealed in Figure \ref{fig:L_eta}, when $L$ reaches its minimum,
the GCR starts to evolve superlinearly; we define the
moment of minimum $L$ as the runaway accretion time $t_{\rm run}$.
For our fiducial model, $t_{\rm run}\simeq$ 10.5 Myrs.
After $t_{\rm run}$, the luminosity grows
as the self-gravity of the envelope becomes
increasingly important. The rise in $L$, together with the relative
constancy of the change in total energy $\Delta E$ (data not shown;
see equations \ref{eq:dt}--\ref{eq:E}),
causes the planet to cool at an ever faster rate, accelerating
the increase in the GCR.

Figure \ref{fig:dt_contr} illustrates the relative importance of 
various terms
in the calculation of time steps
between snapshots (equation \ref{eq:dt}). 
At least until $t_{\rm run}$, the evolution is completely
controlled by changes in the total energy $\Delta E$.
The boundary terms $\langle e_M
\rangle \Delta M$ and $\langle P \rangle \Delta V_{\langle M \rangle}$
are 10--100 times smaller. 

\subsubsection{Understanding our results to order-of-magnitude}
\label{sssec:oom}
\begin{figure}[!htb]
\centering
\includegraphics[width=0.5\textwidth]{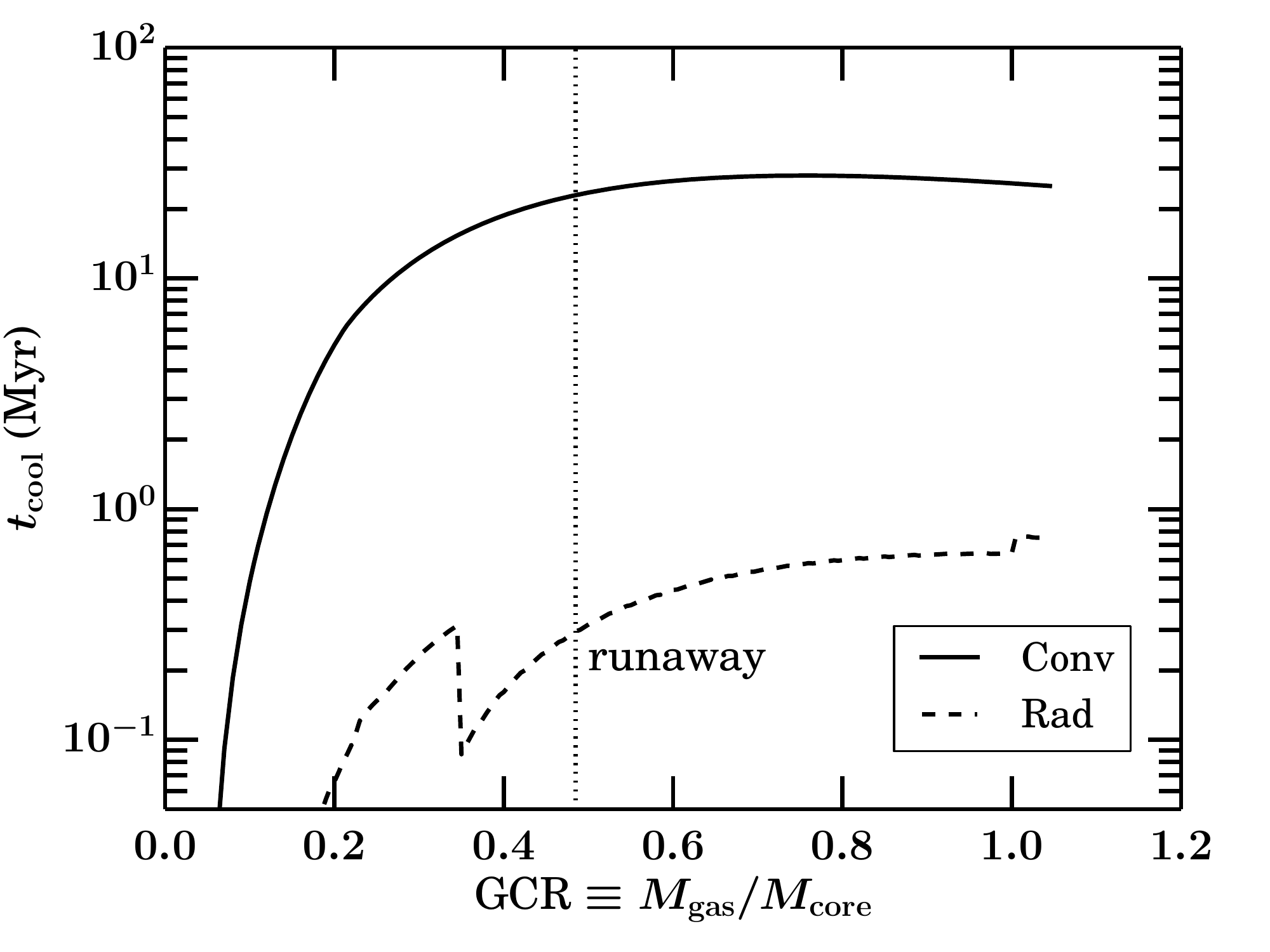}
\caption{\label{fig:tcool_rad_conv} Cooling times of convective (solid
  line) and
  radiative (dashed line) zones for our fiducial model. Since the atmosphere can be
  composed of multiple convective and radiative zones, we plot the
  maximum $t_{\rm cool}$ for each case. The maximum
  convective $t_{\rm cool}$ is always measured in the innermost
  convective zone; it exceeds $t_{\rm cool}$ of any other convective
  zone by 1--4 orders of magnitude. The cooling time in the
  radiative zone abruptly decreases at GCR $\simeq$ 0.35 and increases
  at GCR $\simeq$ 1.0, coinciding with the disappearance and
  re-emergence of an outer convective zone, respectively. The dotted vertical line marks the
  GCR at the moment of runaway. The cooling time of the entire atmosphere
  is dominated by the innermost convective zone, helping to justify
  our assumption of a spatially constant luminosity.}
\end{figure}

The runaway timescale can be approximated as the thermal
relaxation (a.k.a.~cooling) time $t_{\rm cool}$ in the innermost
convective zone, evaluated at GCR $\simeq$ 0.5, a value large enough for
self-gravity to be significant. We define the cooling time of any
zone as its total
energy content divided by the luminosity:
\begin{equation}
t_{\rm cool} = \frac{|E|}{L}\,.
\label{eq:tau_cool}
\end{equation}
Figure \ref{fig:tcool_rad_conv} uses our numerical model 
to evaluate $t_{\rm cool}$ for both convective and radiative zones.
At the moment of runaway,
$t_{\rm cool}$ of the innermost convective zone is $\sim$20 Myr, within a
factor of 2 of $t_{\rm run}\simeq$ 10.5 Myr.

We can also develop back-of-the-envelope understandings of 
$|E|$ and $L$. Our
envelopes are in approximate virial
equilibrium:\footnote{In a strict sense,
  our atmospheres are not isolated objects in virial equilibrium
  because they overlie rocky cores which supply external gravity fields,
  and because their outer boundaries have non-zero pressure and accrete mass.
  These complications generate order-unity corrections to $|E|$;
  at runaway, the atmosphere mass is comparable to the core mass,
  and from Figure \ref{fig:dt_contr} we see that
  the outer boundary terms are at most competitive
  with the total energy.}
the total energy of the atmosphere, of mass
${\rm GCR} \times M_{\rm core}$ and characteristic
radius $R_{\rm RCB}$,
is on the order of the (absolute value of the)
gravitational potential energy:
\begin{equation}
|E| \sim \frac{G M_{\rm core} \times {\rm GCR} \times M_{\rm core}}{R_{\rm RCB}} \,.
\end{equation}
For $R_{\rm RCB} \sim R_{\rm Hill}/3 \sim 15 R_\oplus$, $M_{\rm core} = 5 M_\oplus$,
and ${\rm GCR} \sim 0.5$, we have $|E| \sim 10^{39}$ erg, which is about
1/5 its actual numerically computed value at runaway.
As for $L$,
we know from equation (\ref{eq:grad_rad}) that at the RCB,
\begin{equation}
L = \left. \frac{64\pi GM\sigma T^3\mu m_{\rm H}}{3k \rho \kappa}\nabla_{\rm
  ad} \right|_{\rm RCB} \, .
\label{eq:Lrcb}
\end{equation}
The RCB is always (for dusty models) located at the ${\rm H_2}$
dissociation front where $M_{\rm RCB} \simeq
(1+{\rm GCR})M_{\rm core}$,
$\nabla_{\rm ad} \simeq 0.2$, $T
\simeq 2500$ K, 
and $\kappa \sim 0.1$ cm$^2$/g from H$^-$ opacity.\footnote{The H$^-$
opacity can be fitted by
$\kappa \simeq 10^{-25}\rho^{0.53}T^{7.5}(Z/0.02)$ where all quantities
are evaluated in cgs units.}
A crude estimate of the density at the RCB
is given by ${\rm GCR} \times M_{\rm core} \sim 4\pi R_{\rm RCB}^3
\rho_{\rm RCB}$, where we again take $R_{\rm RCB} \sim R_{\rm H}/3$.
Putting it all together for our fiducial model at ${\rm GCR} \sim 0.5$,
the luminosity thus estimated is $L \sim 10^{25}$ erg/s,
within a factor of 2 of our numerically computed (minimum) value
of $7 \times 10^{24}$ erg/s at runaway.\footnote{The largest
uncertainty in our
order-of-magnitude
calculation is in our estimate of $R_{\rm RCB}$
and by extension $\rho_{\rm RCB}$. Piso \& Youdin (2014)
present a more careful analytic calculation of $L$ at the RCB 
that assumes the outer layer 
is isothermal; for our models, it is generally not.}

\subsubsection{Checks}
\label{sssec:checks}
Our calculation assumes $L$ to be spatially constant---specifically we
assume that the luminosity of the envelope is generated entirely
within the innermost convective zone. To check the validity of this
assumption, we perform a couple {\it a posteriori} tests. We
estimate whether the luminosity generated in radiative zones is
small compared to $L$ and check that most of the planet's thermal
energy content is in the innermost convective zone. From energy
conservation, the luminosity generated in radiative zones that our
model neglects is
\begin{equation}
L_{\rm negl} = -\int_{\rm rad} \rho
  T\frac{\Delta S_M}{\Delta t} 4\pi r^2 dr\,.
\label{eq:dLdr}
\end{equation}
Here $\Delta S_M$ is the difference, taken between snapshots separated
by time $\Delta t$, of $S$ evaluated at the surface enclosing a given
mass $M$. The integral spans all radiative zones. As demonstrated in Figure \ref{fig:dt_contr},
$L_{\rm negl}$ is approximately two orders of magnitude smaller than
the total $L$. In addition, from Figure \ref{fig:tcool_rad_conv} we see that
$t_{\rm cool}$ for the innermost convective zone exceeds $t_{\rm
  cool}$ for any radiative zone or exterior convective zone by at
least an order of magnitude. Because $L$ is constant in our model, an
equivalent statement is that the thermal energy content of the
innermost convective zone exceeds that of any other zone.

\begin{figure*}[!htb]
\centering
\includegraphics[width=\textwidth]{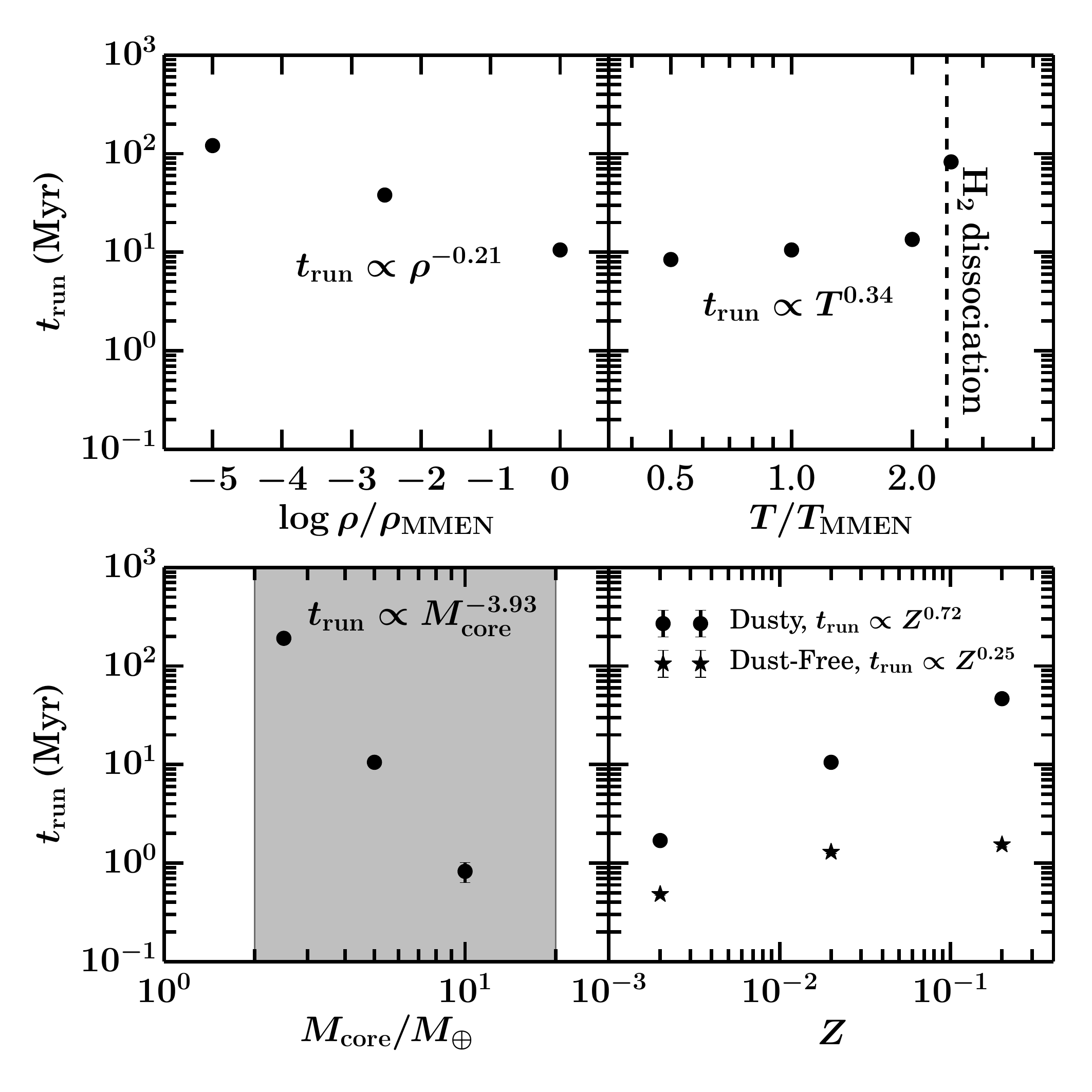}
\caption{\label{fig:trun_vs_param} Runaway time $t_{\rm run}$
  vs.~various boundary conditions: nebular density (upper left),
  nebular temperature (upper right), core mass (lower left), gas
  metallicity and the existence of grains (lower right). 
  Core mass and
  disk metallicity are the most important determinants of $t_{\rm run}$. The gray shaded
region in the lower left panel delineates the observed mass range of
super-Earths from \citet{2014ApJ...783L...6W}. Error bars have
magnitude $\pm t_0$, reflecting our uncertainty in the time of the
first snapshot.
Envelopes become fully convective for $\rho/\rho_{\rm MMEN} \gtrsim 5$
and therefore cannot be evolved in time according to the PY procedure. Nevertheless, 
we surmise that such atmospheres readily run away: their GCRs are on the order of 
unity and their cooling times are short because of the large luminosities required 
to support full-on convection.
}
\end{figure*}

Our results are robust against other shortcomings of our model: 
(1) Can a spatially varying $L$ deepen the RCB so
that the thermal energy content of radiative zones exceeds that of the
innermost convective zone? 
No: in reality, the luminosity has to rise toward the surface; higher
$L$ steepens the radiative temperature gradient (equation \ref{eq:grad_rad}), 
promoting convection in the outer atmosphere and thus lessening the extent
of radiative zones.
(2) What about quantum mechanical effects not captured by our ideal gas EOS?
At high densities ($\rho \gtrsim 0.1 \,\rhocgs$), the mean particle
spacing becomes smaller than the Bohr radius, leading to liquefaction and pressure ionization.
But we find that these packed conditions occur only for
GCR $\gtrsim$ 0.5 (at the moment of runaway)
and only within the bottommost $\sim$3\%
of the planet's atmosphere in radius. 
(3) Finally, how safe is our neglect of the solid core's contributions
to the energy budget of the atmosphere?  The luminosity from
radioactive heating for a $5 M_\oplus$ core is about $8\times 10^{21}$
erg/s \citep[][their Figure 3]{2014ApJ...792....1L}, well below the envelope
luminosities of $\sim$$10^{25}$ erg/s characterizing our fiducial
model (Figure \ref{fig:L_eta}). Whether the core's heat capacity is
significant can be assessed as follows. Over the course of our
atmosphere's evolution, a total energy $\sim$$L t_{\rm run} \sim
10^{25} \erg/{\rm s} \times 10\,{\rm Myr} \sim 3\times 10^{39}$ erg is
released. For the core to matter energetically (either as a source or
sink), it would have to change its temperature by $\Delta T > L t_{\rm
  run} / (M_{\rm core} C_V) \sim 10^4$ K, where $C_V \simeq 10^7
\erg/{\rm K}/{\rm g}$ is the specific heat of rock. Such temperature
changes seem unrealistically extreme, especially over the timescales of 
interest to us---10 Myr---which 
may be short
in the context of solid core thermodynamics.  
For comparison,
some
models of
rocky, convecting super-Earth cores are initialized with temperatures of
5000--20000 K and cool in vacuum over timescales ranging from 0.1--10
Gyr \citep{2012ApJ...748...41S}. Even if by some 
catastrophically
efficient mechanism the core were to lose its entire thermal energy
content over $\sim$10 Myr (say because the viscosity is actually
much lower than that calculated by \citealt{2012ApJ...748...41S};
see, e.g.,
\citealt{papuc08} and \citealt{karato11}), the resultant core luminosity would add to
the envelope contraction luminosity by only a factor of order unity.
\citet[][see their equation 4]{2012ApJ...753...66I} 
typically invoke a core luminosity
of $\sim$$10^{25}$ erg/s --- comparable to our envelope luminosities --- but
only by assuming the entire core can respond thermally
on timescales of $\sim$0.1 Myr (their $\tau_d$).
Such a thermal response time is unrealistically short.

\subsection{Parameter Study}
\label{ssec:dep_on_bc}

We explore how the runaway time changes with various input parameters. Figure
\ref{fig:trun_vs_param} summarizes our results: $t_{\rm run}$ is most
sensitive to core mass and metallicity,
and is insensitive, for the most part,
to nebular density and temperature.

We can understand all of these dependencies as simple consequences of
the properties of the innermost radiative-convective boundary (RCB).
As argued in \S\ref{sssec:oom}, $t_{\rm run}$ is
approximately the cooling time $t_{\rm cool}$ of the innermost
convective zone:
\begin{equation}
t_{\rm run} \sim \left. t_{\rm cool} \right|_{\rm RCB} = \frac{|E|}{L}
\propto \frac{MT}{MT^4\nabla_{\rm ad}/\kappa P} \propto
\frac{\rho\kappa}{T^2\nabla_{\rm ad}}
\label{eq:trun_scale}
\end{equation}
where we have scaled $|E| \propto MT$, and
$\nabla_{\rm ad} \simeq 0.2$ and $T \simeq$ 2500 K 
because the RCB always (for dusty models)
coincides with the H$_2$ dissociation front.
Thus the
variation of $t_{\rm run}$ with input parameters
can be rationalized in terms of $\rho$ and $\kappa$ in equation
(\ref{eq:trun_scale}), as we explain qualitatively below.

\subsubsection{Disk density}
\label{sssec:density}
As Figure \ref{fig:trun_vs_param} attests, $t_{\rm run}$ hardly varies
with nebular density $\rho(R_{\rm out})$. This is because conditions
at the RCB are largely insensitive to nebular parameters, insofar as a 
radiative atmosphere---whose density profile is exponentially
steep---lies between the RCB and the nebula. In a sense, nebular
conditions are increasingly forgotten as one descends toward the RCB
(as found previously by \citealt{1982P&SS...30..755S}).
What little memory remains of outer boundary conditions manifests itself as
a modest increase in $t_{\rm run}$ with decreasing $\rho(R_{\rm
  out})$. At a fixed GCR of 0.5 (characteristic of
runaway), lowering $\rho(R_{\rm out})$ must raise slightly the
interior density, including the density at the RCB, and by extension
the local opacity (which scales as $\rho^{0.53}$).  These small
increments in $\rho_{\rm RCB}$ and $\kappa_{\rm RCB}$ lengthen $t_{\rm
  run}$ according to equation (\ref{eq:trun_scale}). 

\subsubsection{Disk temperature}
\label{sssec:temperature}
The runaway time does not vary much with $T(R_{\rm out}) < 2500$ K for largely
the same reason that it is not sensitive to $\rho(R_{\rm out})$: as pressure
and density e-fold many times across the quasi-radiative outer envelope,
conditions at the RCB decouple from those at the surface.
No matter the value of $T(R_{\rm out}) < 2500$ K, the atmosphere
near the RCB eventually thermostats itself to the H$_2$ dissociation
temperature of 2500 K (cf.~Figure~\ref{fig:atm_prof}, bottom left panel),
and concomitantly strong density gradients serve to isolate the RCB
from the nebula.

Generally, $t_{\rm run}$ increases with disk temperature. As
$T(R_{\rm out})$ approaches the ${\rm H_2}$ dissociation temperature, the outer
envelope becomes increasingly isothermal; the density profile steepens
and $\rho$ and $\kappa$ rise at the RCB. What is
impressive
is the magnitude of the jump in $\rho_{\rm RCB}$, and by extension
$t_{\rm run}$, when $T(R_{\rm out})$
reaches the ${\rm H_2}$ dissociation temperature of $\sim$2500 K and the
outer atmosphere becomes strictly isothermal.
For protoplanetary disks to actually be as hot as $T(R_{\rm out}) \simeq
2500$ K seems unrealistic, since dust sublimation throttles
nebular temperatures to stay below $\sim$2000 K \citep[see, e.g.,][]{d'alessio98,d'alessio01}.

\subsubsection{Core mass}
\label{ssec:Mcore}
Atmospheres atop more massive cores require larger pressure gradients
to maintain hydrostatic support. Increased pressure steepens
radiative gradients (equation \ref{eq:grad_rad}), fostering convection
and pushing the RCB toward the surface (but with $T_{\rm RCB}$ fixed
at $\sim$2500 K). Numerically, we find
$\rho_{\rm RCB}$ decreases by a factor of $\sim$100 from $M_{\rm
  core}=2.5M_{\oplus}$ to 10$M_{\oplus}$; i.e., $\rho_{\rm RCB}
\propto M_{\rm core}^{-3}$. Since $t_{\rm cool}\propto\rho\kappa
\propto \rho^{1.53}$, it follows that $t_{\rm cool}\propto M_{\rm
  core}^{-4.5}$, in rough agreement with the scaling shown in Figure
\ref{fig:trun_vs_param}.

\subsubsection{Opacity: Metallicity and grains}
\label{ssec:opacity}
More metals increase $\kappa$ everywhere,
including at the RCB, where the increased optical thickness 
prolongs $t_{\rm run}$ by reducing the
radiative luminosity.  
Core-nucleated instability is harder at higher
$\kappa$ \citep[e.g.,][]{1982P&SS...30..755S, 2000ApJ...537.1013I,
  2014ApJ...786...21P}. According to our equation
(\ref{eq:trun_scale}), $t_{\rm run} \propto \kappa$; Figure 3 of PY vouches for
this linear dependence. In turn, $\kappa$ scales roughly linearly with metallicity $Z$
in the opacity model by Ferguson et al.~(2005) that we use.

Dusty atmospheres behave differently from dust-free atmospheres.  In
dusty atmospheres, radiative windows opened by dust evaporation
inhibit convection and force the innermost RCB to depths below the
dust sublimation front. In dust-free atmospheres, this impediment to
convection does not exist and so the RCB is free to be located at higher altitudes
where $\kappa$ is smaller. As shown in Figure
\ref{fig:trun_vs_param}, one consequence is that runaway times of
dust-free atmospheres are generally shorter than for dusty
atmospheres. Another consequence is that $t_{\rm run}$ depends only
weakly on metallicity in dust-free atmospheres.  Dust-free envelopes
that are more metal-rich have higher opacities which drive the RCB
outward.  The shortening of $t_{\rm run}$ from decreasing RCB density 
counteracts the lengthening of $t_{\rm run}$ from increasing
opacity.

Note that the trends identified above can reverse if the metallicity 
becomes too high. For $Z\gtrsim 0.5$, increases in the mean molecular 
weight become significant, collapsing the atmosphere and shortening 
$t_{\rm run}$ \citep{2011MNRAS.416.1419H}.

\subsubsection{Outer radius}
\label{sssec:Rout}
It is customary in this field to choose
$R_{\rm out} = \min (R_{\rm H}, R_{\rm B})$.
But the true outer radius may differ from this choice, if only because
there are order-unity coefficients that we have neglected
in our evaluation of $R_{\rm H}$ and $R_{\rm B}$.
How the protoplanet's atmosphere
interfaces with the disk is not well understood. \citet{2009Icar..199..338L}
employ 3D hydrodynamic simulations of planets embedded in viscous disks
to argue that $R_{\rm out}$ should range between $R_{\rm H}/4$
and $R_{\rm B}$ (their equation 3). 
In 2D hydrodynamic simulations, \citet{2014IAUS..299..173O} and 
\citet{ormel14a} vouch for the relevance of $R_{\rm B}$ (when
$R_{\rm B} \ll R_{\rm H}$) to within factors of order unity. Precise
correction factors should depend on the thermodynamic properties
(read: cooling efficiencies) and turbulent/viscous behavior of disk gas
in the vicinity of the planet.

For our fiducial model at 0.1 AU, we find numerically that
$t_{\rm run} \propto R_{\rm out}^{-1.2}$ as $R_{\rm out}$ varies
from $0.5 R_{\rm H}$ to $R_{\rm H}$.
All other factors being equal, larger (i.e., puffier)
atmospheres have lower densities; the lower
value of $\rho \kappa \propto \rho^{1.53}$ at the RCB shortens $t_{\rm run}$
according to equation (\ref{eq:trun_scale}).
In subsequent sections of this paper, we will quote ranges (``error bars'') for $t_{\rm run}$
corresponding to $0.5$--$1 \times R_{\rm out}$. 
Because $R_{\rm H}$ is a hard upper limit on the extent
of planetary atmospheres,
our results for $t_{\rm run}$ when $R_{\rm out} = R_{\rm H}$ (i.e., at orbital
distances $a \lesssim 1$ AU) are hard lower limits.

\section{DISCUSSION: \\HOW DO SUPER-EARTHS GET THEIR GAS?}
\label{sec:disc}

Absent heat sources, gaseous envelopes overlying rocky cores
cool and contract, accreting more gas from their natal disks.
Once these atmospheres become self-gravitating --- i.e., once their
masses become comparable to those of their cores --- they acquire mass
at an accelerating, ``runaway'' rate, ultimately spawning Jovian-class
giants in disks with adequate gas reservoirs. 

How do super-Earths avoid this fate? Perhaps we should revisit
  our assumption of zero heating from planetesimal accretion.  We
  showed in \S\ref{ssec:battery} that in-situ accretion of solids at
  $\sim$0.1 AU finishes well within gas disk lifetimes.  Thus our
  assumption of zero planetesimal accretion seems safe with respect to
  the reservoir of solids that are local to the inner disk. But what
  about solids transported to the inner disk from the outer disk,
  originating from distances $\gg 0.1$ AU?
  Might a steady supply of inwardly drifting planetesimals heat
  super-Earth atmospheres and prevent them from cooling and
  collapsing?
  We can use our results to show that this
  possibility is unlikely. Our model of a $10 M_\oplus$
  core + solar-metallicity envelope under standard nebular conditions
  has a minimum, pre-runaway cooling luminosity of
  $L\sim 5\times 10^{26}$ erg/s. For planetesimal accretion to 
  support this atmosphere against collapse, the planet
  would have to accrete solids at a rate $\dot{M}_{\rm core} \sim
  LR_{\rm core}/(GM_{\rm core}) \sim 1 M_{\oplus}/{\rm Myr}$. Even if
  such an accretion rate could be arranged --- and it would
  require some fine-tuning of planetesimal sizes to get the
  right aerodynamic drift rates 
and accretion efficiencies
 --- sustaining it over
  the $\sim$10 Myr lifetime of the gas disk would cause the
  core to double in mass to $\sim$$20 M_\oplus$. Not only would this
  mass doubling put the planet outside the range of most of the
  measured masses of super-Earths
  \citep{2013ApJ...772...74W, 2014ApJ...783L...6W}, but it would
  also reduce the time to runaway by a factor of $\sim$16
  ($t_{\rm run} \propto M_{\rm core}^{-3.93}$ according to our Figure
  \ref{fig:trun_vs_param}) --- ironically pushing the planet over the
  cliff we were trying to avoid in the first place.
  Thus appealing to accretion of planetesimals, either from the
  inner disk or from the outer disk, to support
  atmospheres against collapse seems infeasible.

\begin{figure*}[!hbt]
\plotone{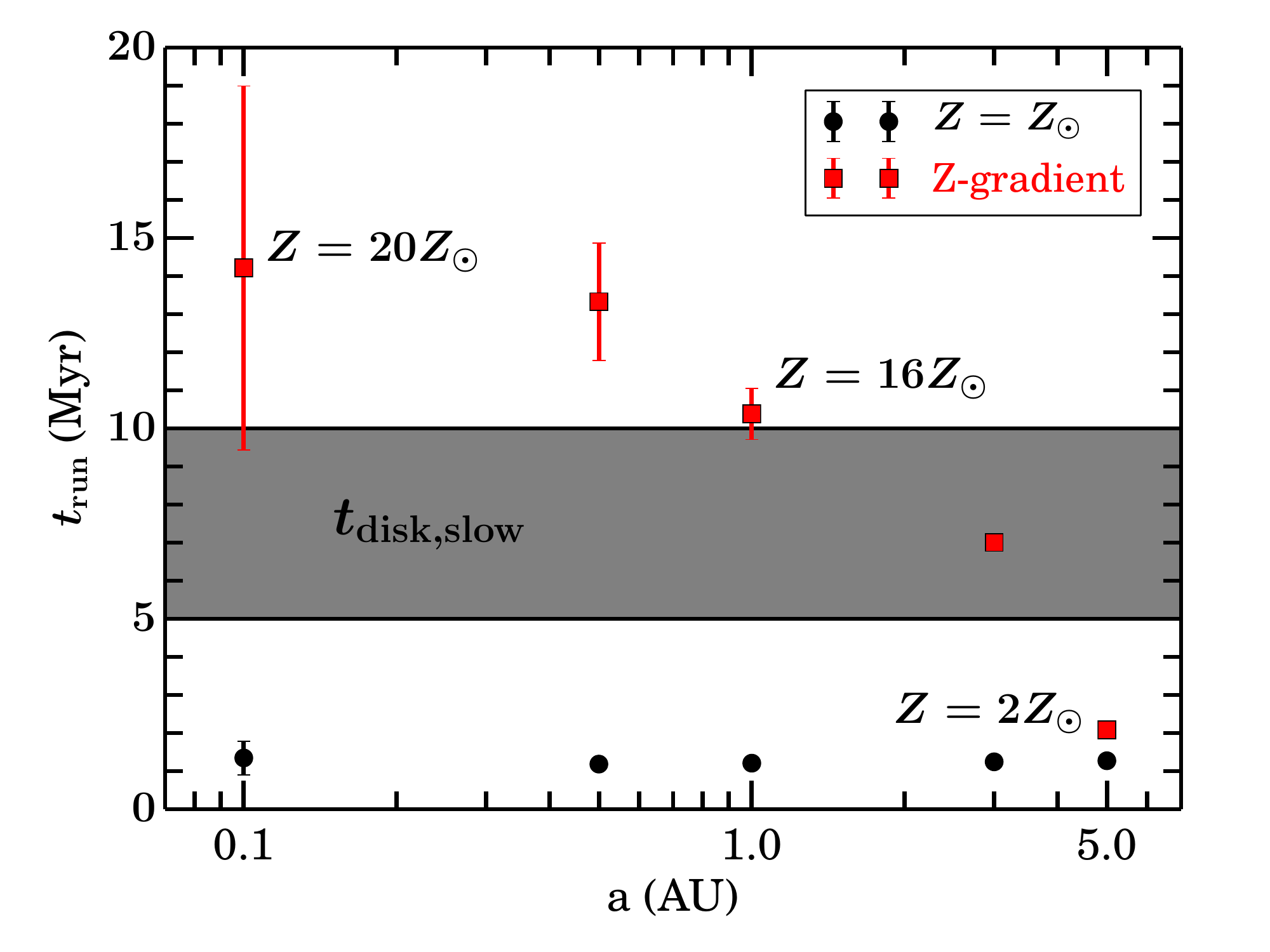}
\caption{\label{fig:trun_vs_a} \scriptsize Runaway time $t_{\rm run}$
  vs.~orbital radius $a$ for 10-$M_{\oplus}$ cores in a dusty disk with
  fixed solar metallicity (black circles) and a dusty disk whose metallicity
  decreases outward but is supersolar throughout (red squares). The
  metallicity trend is assumed linear with $a$. 
 For each $a$, we compute $t_{\rm run}$ at $Z=0.02$
  and $Z=0.2$ and fit a power-law relation for $t_{\rm run}(Z)$; this
  power law (specific to $a$)
  is used to evaluate $t_{\rm run}$ for the desired $Z$. The
  error bars reflect our uncertainty in the atmosphere's outer
  boundary radius, where the upper limit corresponds to $R_{\rm out} =
  0.5 R_{\rm H}$ and the lower limit corresponds to $R_{\rm out} =
  R_{\rm H}$ (see \S\ref{sssec:Rout}); for $a \geq$ 1 AU,
  $R_{\rm B}$ is used instead of $R_{\rm H}$. For disk density and temperature, 
we adopt the MMEN (\S\ref{sssec:BC}), except at $a=0.1$ AU where we take
  $T=2000$ K, since the atmosphere becomes fully convective for $R_{\rm
    out} = 0.5 R_{\rm H}$ and $T=1000$ K and cannot be integrated
  forward. The gray box marks a range in disk lifetimes of 5--10 Myr
  (\citealt{2009AIPC.1158....3M}; 
  \citealt{pfalzner14}). At all orbital distances,
  10-$M_{\oplus}$ cores in a constant solar metallicity disk become
  gas giants within $\sim$1 Myr, well before disk gas
  disperses. By itself, this result (black circles) cannot explain the
  abundance of 10~$M_\oplus$ rocky cores at
  0.05--0.2 AU and the concomitant absence of Jovian-class planets. We
  propose instead that 10~$M_\oplus$ rocky cores coagulated within a
  supersolar disk (red squares): one where metal
  abundances (read: dust opacities)
  are so enhanced at $a \sim 0.1$ AU that
  runaway accretion cannot occur there before the disk gas clears, but
  also where the degree of metal enrichment at $a \sim 5$ AU is
  sufficiently mild to allow the formation of Jupiter.}
\end{figure*}

\begin{figure}[!htb]
\centering
\includegraphics[width=0.5\textwidth]{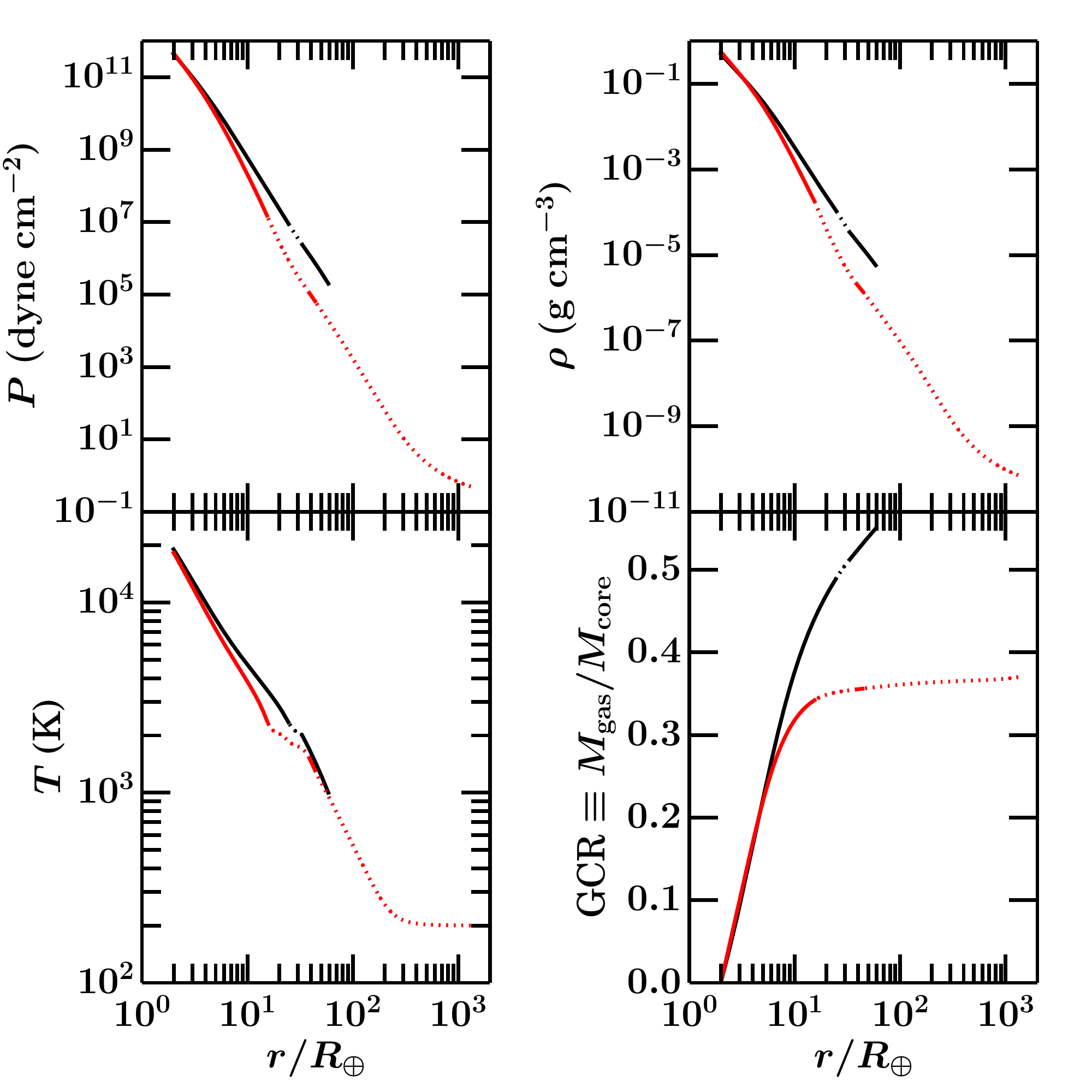}
\caption{\label{fig:orbital_radii_param} Atmospheric profiles 
  just before runaway for 10-$M_{\oplus}$ cores
  at 0.1 AU (black) and 5 AU (red). Dotted lines trace
  radiative zones while solid lines trace convective zones. 
  The innermost convective zone looks practically identical
  between the two models at 0.1 AU and 5 AU; all atmospheric quantities
  in the convective zones agree to within a factor of 2. The similarity
  arises because the radiative-convective boundary which caps the convective
  zone is always (for dusty models) located where H$_2$ first dissociates, and the
  characteristic temperatures and densities required for dissociation
  do not respect orbital location. Consequently, envelope cooling
  times and by extension runaway accretion times are nearly the same
  between 0.1 and 5 AU.}
\end{figure}

With zero heating, the time to runaway gas accretion is the time for
the marginally self-gravitating gas envelope to cool.
As the black points in Figure \ref{fig:trun_vs_a} reveal, at a fixed
envelope metallicity and core mass, the runaway time is remarkably invariant
with orbital distance. Figure \ref{fig:orbital_radii_param} shows
why. The radiative-convective boundary (RCB) of the atmosphere 
occurs at the ${\rm H_2}$ dissociation front. Because the
circumstances of ${\rm H_2}$ dissociation are fairly universal
(governed as they are by the universal laws of statistical quantum
mechanics), the H$_2$ dissociation front occurs at temperatures
and densities that are insensitive to 
whether the planet is located at 0.1 AU or 5 AU.
Temperature and density profiles in the convective zone
interior to the RCB vary only by factors of 2 between models at
different stellocentric distances. This inner convective zone
contains the lion's share of the envelope's mass and energy,
which means that
its cooling rate controls the time to runaway accretion.
Nearly identical convective zone profiles
beget nearly identical runaway accretion times.
Similar results are reported by \citet{2011MNRAS.416.1419H}, who
find for their static models
that the critical
core mass --- the maximum core mass for which the envelope can stay in
strict hydrostatic equilibrium --- varies by at most a factor of 2 from 1 AU to 10 AU.
Like us, these authors incorporate ${\rm H_2}$ dissociation in their
equation of state and opacity laws.

Taken at face value, the black points in Figure \ref{fig:trun_vs_a}
suggest that 10-$M_{\oplus}$ solid cores, placed anywhere from 0.1 AU
to 5 AU in a gas-rich protoplanetary disk of solar composition, readily transform into
gas giants before the gas disperses in 5--10 Myr. But observations
inform us that 10~$M_\oplus$ rocky cores---and not gas giants---abound
at distances inside 1 AU. The {\it Kepler}
spacecraft has established that super-Earths having radii of 1--4 $R_{\oplus}$
orbit some $\sim$20\% of Sun-like stars
at distances of 0.05--0.3 AU \citep{2010Sci...330..653H, 2013ApJS..204...24B,
  2013ApJ...770...69P, 2013ApJ...778...53D, 2013ApJ...766...81F, 2014ApJ...784...45R}.
By contrast, Jupiter-sized objects are rare; the 
occurrence rate for hot Jupiters inside $\sim$0.1 AU
is only $\sim$1\%, and
the occurrence rate for warm Jupiters between $\sim$0.1--1 AU 
is even smaller, in the so-called ``period valley'' 
\citep{jones03, udry03, 2012ApJ...753..160W,
2013ApJ...766...81F,  
2013ApJ...767L..24D}. At larger distances, gas giants appear more
frequently, orbiting up to $\sim$20\% of Sun-like stars at $a<10$ AU
\citep{2008PASP..120..531C}. 

How do we reconcile our models with these
observations? We propose two possible scenarios:
(1) cores accrete envelopes in 
disks with dust-to-gas ratios that are strongly
supersolar at 0.1 AU and that decrease outward, and
(2) the final assembly of super-Earths is delayed
by gas dynamical friction to the era of disk dispersal.

\subsection{Supersolar Metallicity Gradients in Dusty Disks}
\label{ssec:Zgradient}

Enriching atmospheres in metals (by increasing their
dust content or, less effectively, by increasing their
metallic gas content)
delays runaway accretion by making envelopes more opaque, decreasing their
luminosities and extending their cooling times (see \S\ref{ssec:dep_on_bc}).
The importance of metallicity and opacity in this regard is widely acknowledged
(e.g., \citealt{1982P&SS...30..755S}; \citealt{2000ApJ...537.1013I}; PY). 
Our goal is to search for an appropriately supersolar and outwardly
decreasing metallicity profile for the parent gas disk that can
prevent gas giant formation at 0.1 AU while promoting it at 5 AU.

Disk metallicity gradients are actually
hinted at by the atmospheric compositions of close-in
super-Earths GJ 1214b (6.26 $M_{\oplus}$, 2.85 $R_{\oplus}$, $a=0.014$
AU; \citealt{2013A&A...549A..10H}) and GJ 436b (24.8 $M_{\oplus}$, 4.14
$R_{\oplus}$, $a=0.030$ AU; \citealt{2012ApJ...753..171V}), and our own
Jupiter. GJ 1214b and GJ 436b are characterized by optical-to-infrared
transmission spectra that are featureless \citep{2014Natur.505...69K,
  2014Natur.505...66K}. Clouds can explain these flat spectra, but the
kinds of clouds that are compatible with observations can only be
generated in atmospheres of supersolar metallicity \citep[e.g., $Z
  \simeq 0.4 \simeq 20 Z_\odot$:][]{2013ApJ...775...33M}. Higher
metallicity envelopes have more condensibles so that cloud formation
occurs at the higher altitudes probed by near-infrared observations.\footnote{Figure 1 of
  \citet{2013ApJ...775...33M} suggests that at $Z\simeq 0.4$,
  the cloudbed forms at
  $\sim$30 mbar, a pressure $\gtrsim 30$ times
  higher than the observationally inferred cloud-top pressure of
  $\lesssim 1$ mbar for GJ 1214b \citep[][their Figure
  3]{2014Natur.505...69K}. \citet{2013ApJ...775...33M} argue that
  the cloud particles (of ZnS and KCl)
  can be lifted by currents or turbulence.
  We note that the need for vertical updrafts 
  lessens as super-Earth atmospheres increase in metal content,
  and in fact, observations are compatible with $Z$ up to 1
  (see also \citealt{moses13}). But we
  disfavor $Z \gtrsim 0.5$ because
  envelopes with such extreme
  metallicity---with their large
  mean molecular weights $\mu$---tend to runaway quickly
  \citep{2011MNRAS.416.1419H, 2014ApJ...786...21P}. 
  As long as $Z \lesssim 0.5$, $\mu$ 
  increases only weakly
  with $Z$ \citep[][their Figure 6]{2011ApJ...733....2N}.}
Jupiter at 5 AU is also observed to have supersolar metallicity,
but importantly, the degree of enrichment is less extreme than for
close-in super-Earths. In-situ measurements of elemental abundances by
the {\it Galileo} probe indicate that Jupiter's upper atmosphere has
$Z \simeq 0.04$ (\citealt{owen99}; see also \citealt{2005AREPS..33..493G}). 
Models of Jupiter's interior using equations of state based on laser
compression experiments suggest that Jupiter's envelope as a whole has
$Z=0.02$--$0.1$ \citep[][their Figure 7]{2005AREPS..33..493G}.

In Figure \ref{fig:trun_vs_a}, we demonstrate that a simple linear metallicity
profile extending from $Z=0.4$ $(20Z_\odot)$ at 0.1 AU to $Z=0.04$
$(2Z_\odot)$ at 5 AU can successfully circumvent runaway at $a \lesssim 1$ AU,
while still ensuring the formation of Jupiter at 5 AU. Strong metal
enrichment at $\sim$0.1 AU protects close-in super-Earths from
becoming gas giants. As one travels down the metallicity gradient, the
time to runaway decreases and eventually falls within gas
disk lifetimes, in accord with the outwardly increasing occurrence
rate of gas giants.
Note that the lengthening of runaway time at $\sim$0.1 AU
is made possible by refractory (silicate/metal) dust grains,
whose evaporation causes the innermost radiative-convective boundary (RCB) to coincide
with the H$_2$ dissociation front. This placement of the RCB 
makes $t_{\rm run}$ especially sensitive to the overall
gas metallicity, since metals contribute to the abundance of
H$^-$ which dominates the local opacity.

A supersolar and outwardly decreasing metallicity profile in the
innermost regions of protoplanetary disks is not without physical
motivation. First note that disk metallicity should not be confused
with host star metallicity. During the earliest stages of star/planet
formation, a protostellar disk may begin with a spatially uniform
metallicity equal to that of its host star. But thereafter, dust and
gas within the disk can segregate, and the metallicity can vary with
location. The protoplanetary disk TW Hydra is observed to have a
dust-to-gas ratio that decreases radially outward
(\citealt{2012ApJ...744..162A}; see also \citealt{williams14}).
This decreasing metallicity profile is
readily explained by solid particles drifting inward by aerodynamic
drag and possibly piling up (\citealt{2002ApJ...580..494Y}; \citealt{2004ApJ...601.1109Y};
\citealt{birnstiel12}; \citealt{2013ApJ...775...53H}; \citealt{2014ApJ...780...53C};
\citealt{schlichting14a}).
In turn, increased solid abundances reduce radial drift velocities,
fostering stronger pile-ups \citep{2010ApJ...722.1437B}. 
The collection of solid material amassed in the inner disk, coupled with
the higher orbital speeds there, enhances local collision rates and
collision velocities. High-speed collisions shatter solids, polluting
the surrounding nebular gas with dust---though whether such dust can
persist in planetary atmospheres and avoid coagulation and
sedimentation is not clear \citep{2014ApJ...789L..18O,
  2014arXiv1406.4127M}. 
Another concern is whether 
icy mantles of drifting
dust grains may have sublimated
away before reaching the hot inner disk;
gas there might then be too poor in volatile
species like C, N, and O to explain
their inferred abundances in super-Earth atmospheres
\citep{2013ApJ...775...33M, 2014Natur.505...66K, 
2014Natur.505...69K}. 
One way out is to imagine that sufficiently large planetesimals
keep their volatiles locked in their interiors
as they drift past the disk's nominal ice line.
Another possibility is that disk gas can accrete and
transport gaseous volatiles created at the sublimation front.

\begin{figure*}[!htb]
\plotone{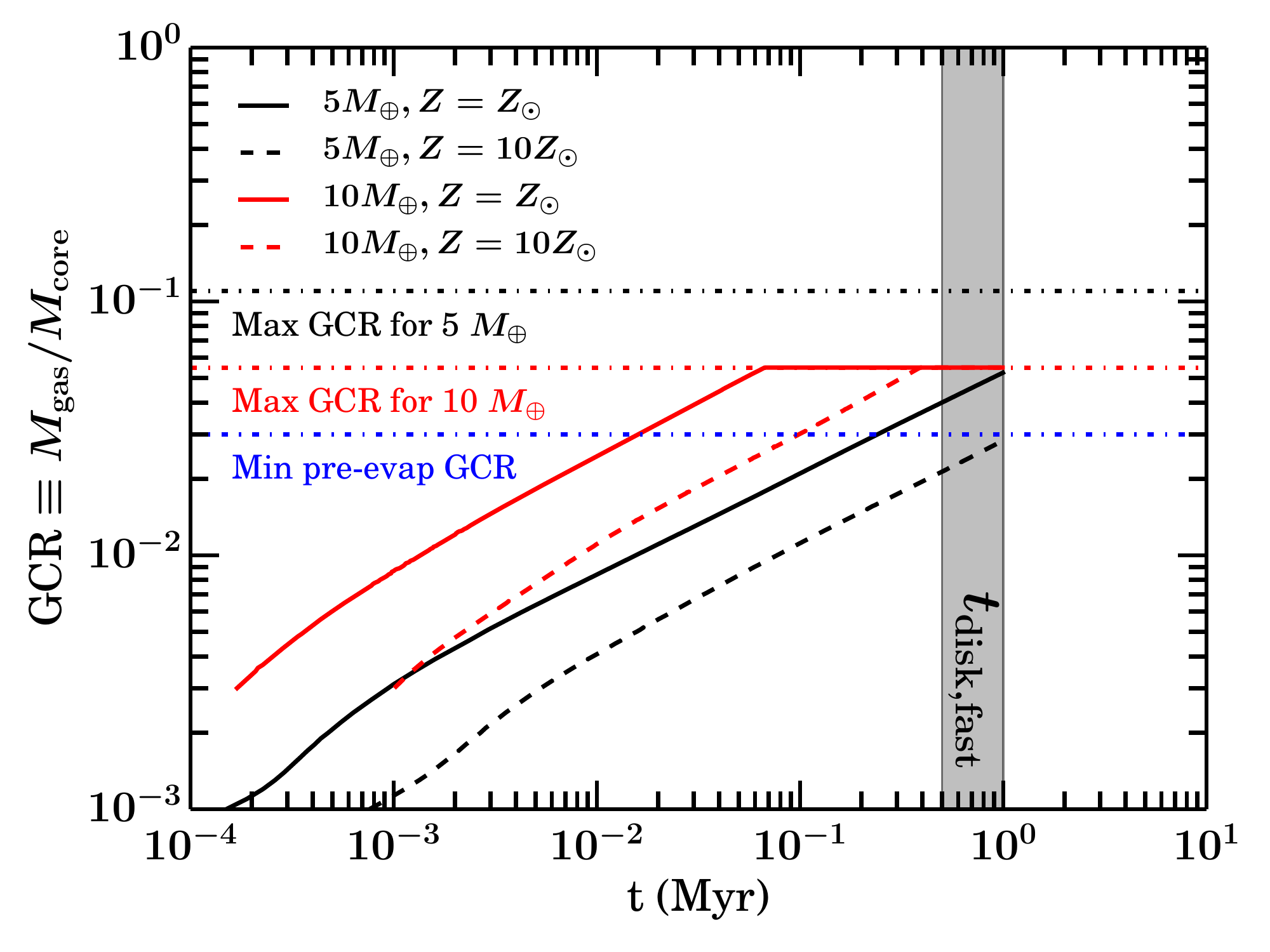}
\caption{\label{fig:rhoreduced}The evolution of gas-to-core mass ratio
(GCR) for 5-$M_{\oplus}$ (black) and 10-$M_{\oplus}$ (red) cores in a
disk with density $\rho = 10^{-3}\rho_{\rm MMEN}$.
Solid/dashed curves
represent solar/supersolar metallicities.
In the modern view of disk dispersal, disks last for $t_{\rm disk,
  slow} \sim$ 5--10 Myr \citep{2009AIPC.1158....3M, pfalzner14}, after which the
gas rapidly dissipates over a timescale $t_{\rm disk, fast} \sim $
0.5--1 Myr \citep{alexander14}.
The gray box denotes $t_{\rm disk, fast}$. 
Dash-dotted lines denote maximum GCRs for
5-$M_{\oplus}$ (black) and 10-$M_{\oplus}$ (red) cores,
assuming that in any given disk,
all the gas inside 0.1 AU is completely and equitably accreted
by three identical cores.
The red curves flatline because gas is completely depleted by accretion 
onto 10-$M_{\oplus}$ cores before disk dispersal. The blue dash-dotted 
line denotes the approximate minimum GCR that planets should have 
before photoevaporation, so that after photoevaporation
the GCR matches observationally inferred values
\citep{2013ApJ...775..105O, 2014ApJ...792....1L}.}
\end{figure*}

The scenario presented in this subsection posits that super-Earth
cores form within disks having full reservoirs of gas.
Although our models indicate that cores in such gas-rich nebulae
stave off runaway for high disk metallicities,
we find that the planets do not avoid accreting fairly massive gas envelopes.
Using our scaling relations, we estimate that
at $a=0.1$ AU in a $Z=0.4$ gas disk that lasts 5--10 Myr,
a 10~$M_\oplus$ core attains a gas-to-core mass ratio
GCR of $\sim$0.2--0.6.\footnote{The GCR as a function of time for
  $Z=0.4$ is calculated by dilating the time axis of GCR$(t)$ for
  $Z=0.2$ by $2^{0.72}$ (see lower right panel of Figure
  \ref{fig:trun_vs_param}).} 
These values (which lie within order-unity factors of runaway) are
considerably higher than present-day GCRs, which apparently range
from $\sim$0.03--0.1
\citep[for {\it Kepler} planets having 2--4 $R_\oplus$;][their Figures
6 and 7]{2014ApJ...792....1L}.
Photoevaporation can bridge the gap between 
past --- i.e., the moment the gas disk clears --- and present.
Over the course of $\sim$100 Myrs, X-rays from 
host stars can photoevaporate super-Earth envelopes 
from initial GCRs of $\sim$0.4 
down to final GCRs of $\sim$0.01--0.1, with
the precise evolution depending on stellocentric distance
and core mass \citep[][their Figure 8]{2013ApJ...775..105O}.
In fact, planets of any initial GCR from $\sim$0.01--0.4, when eroded by X-rays, 
tend to asymptote toward a final GCR of $\sim$0.01.
This convergence arises because higher mass envelopes
at early times are more distended and therefore
lose mass more quickly than lower mass envelopes
at late times: photoevaporative histories that begin differently
conclude similarly.

\subsection{Late-Stage Core Formation in\\Gas-Depleted Nebula}
\label{ssec:depletion}

Another way to prevent gas giant formation on close-in orbits is to delay the
final assembly of cores until the era of disk dispersal. 
It may seem that we will have to fine-tune the timing
of disk dispersal and the degree of nebular density reduction 
so that cores acquire enough gas ($\sim$1--10\% by mass) to satisfy
observations. But we show below that there is a wide
range of acceptable scenarios: that the nebula 
can be reduced in density by factors as large as
$\sim$1000 and still provide enough gas to reproduce the inferred 
atmospheres of super-Earths.

Gas exerts dynamical friction on proto-cores, postponing mergers
by damping eccentricities and preventing orbit crossing.
Deferring the final coagulation of solids until after
the gas clears and dynamical friction weakens
is the standard way to explain how the
terrestrial planets in our solar system
avoided accreting nebular hydrogen \citep[e.g.,][]{2002Icar..157...43K}. 
Though it may not be obvious,
we will see from a timescale comparison given below that even within this scenario
of late-stage core assembly, the assumption of zero power from the accretion of solids
($L_{\rm acc} = 0$) during the era of gas accretion can still be valid at
$\sim$0.1 AU. Gas dynamical friction delays the final merger phase of proto-cores, 
but once this phase begins, it completes rapidly so that subsequent gas accretion occurs 
without solid accretion.

We consider the final assembly of 10~$M_\oplus$ cores
from merging pairs of 5-$M_\oplus$ cores, i.e.,
the last doubling in planet mass that follows after an ``oligarchy''
of multiple 5-$M_\oplus$ proto-cores destabilizes
and crosses orbits \citep[e.g.,][]{1998Icar..131..171K}.
The timescale for gas dynamical friction to damp
the eccentricity of a proto-core (and thereby forestall orbit crossing) at 0.1 AU 
in the minimum-mass extrasolar nebula (MMEN) is 
\begin{equation}
t_{\rm friction} \simeq 0.1\, {\rm yr} \left(\frac{T}{10^3 \, {\rm
      K}}\right)^{3/2}\left(\frac{6\times
    10^{-6}\,\rhocgs}{\rho}\right)\left(\frac{5 \,M_{\oplus}}{M_{\rm
      core}}\right)\, 
\label{eq:tgravdrag}
\end{equation}
(see, e.g., equation 2.2 of \citealt{2002Icar..157...43K}).
This stabilization timescale should be compared against the
destabilization (a.k.a.~viscous stirring) timescale for oligarchs
to cross orbits by mutual gravitational interactions.
Drawing from the viscous stirring formulae of \citet{2004ARA&A..42..549G}, we find that 
orbit crossing occurs when the gas surface density
falls below the surface density of oligarchs.
In other words, when the density of gas becomes comparable
to that of solids --- i.e., when $\rho \sim 5 \times 10^{-3} \rho_{\rm MMEN}$ --- one can no longer treat the gas disk
as an infinite sink of angular momentum for the oligarchs;
the backreaction on gas by the oligarchs effectively
shuts off gas dynamical friction. Oligarchs
proceed to excite each other's eccentricities to the point
of orbit crossing and merging.

Such depleted disks do not have enough gas to spawn gas giants.
But is there enough nebular gas remaining for cores to accrete 
envelopes massive enough to satisfy observationally inferred GCRs?
Our answer is yes, for core masses $\gtrsim 5 M_{\oplus}$
and for gas densities $\rho$ not much less than $10^{-3} \rho_{\rm MMEN}$.
Figure \ref{fig:rhoreduced} shows that 
within a disk whose gas content has drained 1000-fold relative
to that of the MMEN, cores of mass 5--10 $M_\oplus$ can still 
accrete enough gas to attain GCRs of 2--5\% 
within $t_{\rm disk, fast} \sim 0.5$--1 Myr --- this is our estimate for
the timescale over which disk gas exponentially decays.
We compute the latter by taking 10\% of $t_{\rm disk, slow} \sim 5$--10 Myr,
the age of the disk when it first begins
to dissipate.
For a review of the ``two-timescale'' nature of disk dispersal, see
\citet{alexander14}.

The process of orbit crossing and merging,
once begun, completes on a timescale much shorter than 
$t_{\rm disk,fast}$. For 5-$M_{\oplus}$ oligarchs separated by $10$ mutual Hill radii at 0.1 AU,
we estimate that the orbit-crossing timescale $t_{\rm cross}$ 
can range anywhere from $\sim$3 to $\sim$3000 yr,
where we have scaled the results of
\citet[][the squares and crosses in their Figure 1a]{zhou07} 
at 1 AU for the shorter orbital period at 0.1 AU,
and where the range in times reflects a 
possible range of non-zero
initial eccentricities and inclinations.
For our chosen parameters, both $t_{\rm cross}$ and 
the coagulation timescale $t_{\rm coagulate} \sim 10^4$ yrs
(Equation (\ref{eq:tcoag3}) in Section \ref{sec:intro}) are still 
small fractions of $t_{\rm disk, fast}$.
This validates our assumption that planetesimal accretion is negligible
while gas accretes onto cores.
To re-cap the 
sequence of events: (a) oligarchs
are prevented by gas dynamical friction from
completing their last doubling
for $t_{\rm disk, slow} \sim 5$--10 Myr; (b) the gas density depletes by
a factor of $\sim$200
over several
e-folding times $t_{\rm disk, fast} \sim 0.5$--1 Myr until 
the gas density becomes comparable to the solid density
and dynamical friction shuts off; 
(c) neighboring oligarchs perturb one another
onto crossing orbits over $t_{\rm cross} \sim 3$--3000 yr;
(d) super-Earth cores congeal
within $t_{\rm coagulate} \sim 10^4$ yr; at this
stage planetesimals are completely consumed; (e) whatever
nebular gas remains is accreted by super-Earths within the next gas e-folding
timescale $t_{\rm disk,fast} \sim 0.5$--1 Myr --- a phase during which there is
no planetesimal accretion ($L_{\rm acc} = 0$).

Our final GCRs for 10~$M_\oplus$ cores
are $\sim$5--20\%
for disk gas densities $1/1000$--$1/200$ that of the MMEN.
X-ray photoevaporation can whittle our computed GCRs
down to a few percent
or lower \citep{2013ApJ...775..105O}.
Post-evaporation GCRs
of a few percent agree with GCRs estimated from observations of
present-day super-Earths (\citealt{2014ApJ...792....1L}; note that
inferred GCRs can be as low as 0.1\%; see their Table 1).

While super-Earths at 0.1 AU can
achieve GCRs of a few percent,
their counterparts at 1 AU may be devoid of gas. The reason 
is that coagulation times for solids --- in the absence of gravitational focussing ---
increase strongly with orbital distance: $t_{\rm coagulate} \propto a^{3.5}$ 
(see equation \ref{eq:tcoag3}). 
Contrast the sequence of events outlined above for
cores at 0.1 AU with the situation at 1 AU where
$t_{\rm cross} + t_{\rm coagulate} \sim 3 \times 10^7$ yr $\gg t_{\rm disk, fast}$
(where again we have drawn $t_{\rm cross}$ from \citealt{zhou07}).
Thus upon assembly at 1 AU, Earth-sized and larger cores have no gas at all
to accrete. What little gas may have been accrued by proto-cores
before they merge may be blown off 
after they fully coagulate,
by accretion of planetesimals (having sizes $\gtrsim 2$ km for a
$1 M_{\oplus}$ proto-core; \citealt{schlichting14})
or by Jeans escape and hydrodynamic escape
(e.g., \citealt{watson81}; see also the textbook
by \citealt{chamberlain87}).

Even farther out at $\sim$5 AU, the process of core accretion must
perform an abrupt about-face.  Here we desire that cores massive
enough to undergo runaway coagulate within gas-rich disks, in order
that gas giants like Jupiter may form. 
Cores having isolation masses at these distances must run away
without having to merge with neighboring bodies.
The standard argument is to appeal to the boost in isolation masses at larger orbital radii. 
Protoplanets farther out have larger feeding zones, both because
of their larger orbits and because of their larger Hill radii.
Furthermore, the disk's solid surface density is enhanced
outside the ``ice line'' (water condensation front) at $\sim$2--3 AU 
(\citealt{2006ApJ...640.1115L}; see also
\citealt{2011ApJ...743L..16O}). The factor of $\sim$4 increase in
solids from ice condensation raises oligarch masses by a factor of 
$4^{3/2}$ to $\sim$$5 M_\oplus$ at 5 AU within the
minimum-mass solar nebula (MMSN; see, e.g., equation 22 of
\citealt{2000Icar..143...15K}).
These masses are within factors of
2 of those required for runaway gas accretion within gas disk lifetimes.
A modest, order-unity increase in solid surface density above that of the MMSN
(provided, e.g., by the MMEN, whose density exceeds that of the MMSN
by a factor of 5)
can easily make up the shortfall.

Details of the scenario described in this subsection are subject to
some uncertainty. Cores may not accrete sufficient gas if planetesimal 
accretion rates are high 
enough (N. Inamdar \& H. Schlichting, in preparation).
Our scenario also neglects gas disk
turbulence and its associated density fluctuations, which can cause
oligarchs' semi-major axes and eccentricities to random walk
(\citealt[][their section 3.1]{2012ARA&A..50..211K};
\citealt{2013ApJ...771...43O}). Thus gas
does not only delay core
formation through dynamical friction; it can also hasten core
formation by promoting orbit crossing through turbulent stirring. 

\section{CONCLUSIONS}
\label{sec:concl}

Observations and modeling of the radii and masses of close-in
super-Earths reveal that such planets may have hydrogen envelopes
comprising a few percent by mass of their solid cores.  We calculated
how rocky cores could accrete such atmospheres from their natal gas disks,
under a wide variety of nebular conditions and at orbital distances
ranging from 0.1 to 5 AU.
Our main findings are as follows:

\begin{enumerate}
\item In 
an in-situ formation scenario, solids coagulate to form 
close-in super-Earth cores, consuming
  all available planetesimals, well before gas accretes onto those
  cores. High local
surface densities and short dynamical times enable fast
  coagulation.  
The luminosity from planetesimal accretion
likely cannot prevent runaway
for 10~$M_\oplus$ cores,
even when we account for planetesimals
that originate from outside $\sim$1 AU. 
With no planetesimal accretion as a heat source,
  the evolution of the gaseous envelope is
  that of Kelvin-Helmholtz contraction: envelopes gain mass as fast as 
	they can cool.
\item The time $t_{\rm run}$ for a core to undergo runaway gas
  accretion is well approximated by the cooling time of the envelope's
  innermost convective zone.  The extent of this zone 
  is determined by where H$_2$ dissociates --- when envelopes are dusty.
  The strong dependences
  of $t_{\rm run}$ on core mass and dust opacity, and its weak
  dependences on nebular density and temperature, can be understood 
  in terms of the circumstances governing H$_2$ dissociation.
\item In disks with solar metallicity and gas densities comparable to
  either the minimum-mass solar nebula (MMSN) or the minimum-mass
  extrasolar nebula
  (MMEN) --- these differ only by factors of a few in density ---
  10~$M_\oplus$ cores undergo runaway gas accretion to become
  Jupiters, irrespective of whether they are located at 0.1 or 5
  AU. The propensity for super-Earths at $\sim$0.1 AU to explode
  into Jupiters is at odds with the rarity of gas giants at these
  distances.  We presented two ways to solve this puzzle:
	\begin{enumerate}
	\item 
		Disks have gradients in their dust-to-gas ratio. 
		To prevent runaway at distances $< 1$ AU yet ensure the formation of Jupiters
		at $\sim$1--5 AU, the inner disk may have to have a strongly supersolar 
		dust-to-gas ratio (e.g., $Z=0.4=20Z_\odot$), while the outer disk may be more 
		nearly solar in metallicity (e.g., $Z=0.04$). Copious dust 
pushes the boundary of the innermost convective zone to the H$_2$ dissociation
front, where increased metals enhance the H$^-$ opacity and slow cooling.
		We estimated that 10-$M_{\oplus}$ cores at 0.1 AU in dusty $Z=0.4$
		disks achieve gas-to-core mass ratios (GCRs) that are 
    marginally small enough to avoid runaway. 
		After the disk gas clears, high-energy radiation from 
		host stars photoevaporates planetary envelopes and 
		can reduce GCRs to a few percent, in line with observation.

\item Super-Earth cores coagulate just as the gas is about to
  disappear completely. Coagulation is inhibited by gas dynamical
  friction; proto-cores merge to become full-fledged super-Earths only after the gas 
surface density drops below the surface density of proto-cores 
so that dynamical friction shuts off. A 10~$M_\oplus$ core that forms in such a depleted 
nebula at 0.1 AU can still emerge with
a GCR of a few percent or larger, even after photoevaporative erosion.
	\end{enumerate}

Note that scenarios (a) and (b) are not mutually exclusive. In fact, (b) can reinforce (a): 
solids can be left behind while disk gas depletes (say by disk photoevaporation; e.g.,
\citealt{guillot06}), increasing 
dust-to-gas ratios.

\end{enumerate}

One way to test our ideas
is to measure the occurrence rates of
super-Earths and Jupiters as functions of orbital distance ---
particularly beyond $\sim$1 AU.  Our expectation is that super-Earths
should become less common at distances $\gtrsim 1$ AU as they are transformed
into Jupiters.
The frequency of Jupiters is already known to increase outward
--- from the ``period valley'' at $\sim$0.1 AU where giant planets are
rare, to the ``land of the giants'' at $\sim$1--10 AU where occurrence
rates can be as large as $\sim$20\% \citep{2008PASP..120..531C}.  The
data for super-Earths is less extensive; occurrence rates for objects
having radii $1.25$--$2 R_{\oplus}$ per logarithmic bin in orbital period
are nearly uniform out to 145 days \citep{2013ApJ...766...81F}.  But
for objects with $2$--$4 R_\oplus$ (what we still categorize
as super-Earths but which \citealt{2013ApJ...766...81F}
call ``mini-Neptunes''), there are exciting hints
that the occurrence rate decreases with increasing distance: their
frequency decreases monotonically from $\sim$6\% at $\sim$30 days to
$\sim$3\% at $\sim$150 days \citep{2013ApJ...766...81F}. 
We look forward to extending this data to
still longer periods with radial velocity and transit surveys.

To further evaluate the disk metallicity gradient scenario, it will be helpful
to measure dust-to-gas ratios against orbital distance in disks
(e.g., \citealt{2012ApJ...744..162A}; \citealt{williams14}),
especially inside a few AU, and
to constrain metallicities of more super-Earth atmospheres, as was
done for GJ 436b \citep{2014Natur.505...66K} and GJ 1214b
\citep{2014Natur.505...69K}.  On the theoretical side, the supersolar
metallicity scenario requires that atmospheres be dusty, but efficient
coagulation and sedimentation can clear atmospheres of dust; thus,
more studies of cloud/grain physics will also be welcome
\citep{2013ApJ...775...33M, 2014ApJ...789L..18O, 2014arXiv1406.4127M}.
More generally, our
1D models can be improved by considering 2D/3D effects such as the
opening of gaps in circumstellar disks \citep[e.g.,][]{fung14} and the formation of
circumplanetary disks (\citealt{2009Icar..199..338L}; \citealt{ormel14a}). 
And more carefully resolving the outermost
radiative zones of our model atmospheres will enable us to explore a
wider range of core masses and outer envelope radii.

There are a couple ``extreme solar systems'' deserving of further consideration.  
Kepler-36 hosts two planets that are only 0.01 AU apart yet whose
measured densities suggest one is purely rocky while the other contains
significant gas \citep{carter12}. 
How can this system be
accommodated within core accretion theory (see
\citealt{2013ApJ...775..105O} and \citealt{lopez13} for pioneering
explorations)?  HD 149026b is a ``hot Saturn'' situated at 0.04 AU
with a total mass of 114 $M_{\oplus}$ and a modelled core mass of
$\sim$$67 M_{\oplus}$ \citep{sato05,wolf07}: the latter is high enough for the
planet to become a Jupiter at any orbital distance.  How did it avoid
becoming a gas giant?

\acknowledgments
We are indebted to Jason Ferguson for extending and
sharing his opacity tables.
We thank Rebekah Dawson, Jonathan Fortney, Brad Hansen, Howard Isaacson, Doug Lin, Eric Lopez,
Geoff Marcy, Burkhard Militzer, Tushar Mittal,
Ruth Murray-Clay, Erika Nesvold, James Owen, Alex Parker, Erik Petigura, Roman Rafikov,
Hilke Schlichting,
Jonathan Williams, Yanqin Wu,
and Andrew Youdin for helpful discussions.
We are grateful to an anonymous referee for providing a thoughtful
and encouraging report.
EJL is supported in part by the
Natural Sciences and Engineering Research Council
of Canada under PGS D3 and the Berkeley Fellowship.
EC acknowledges support from a Berkeley Miller Professorship,
and grants AST-0909210 and AST-1411954
awarded by the National Science Foundation.
CWO acknowledges support 
from the National Aeronautics and Space Administration (NASA)
through Hubble Fellowship grant HST-HF-51294.01-A
awarded by the Space Telescope Science Institute, which is
operated by the Association of Universities for Research in
Astronomy, Inc., for NASA, under contract NAS 5-26555.

\bibliography{superearth}

\end{document}